\documentclass[a4paper,11pt]{article}
\usepackage{jcappub} 
\usepackage{lineno}
\usepackage{color,amsmath,amssymb,graphicx,latexsym,subfigure}
\pdfoutput=1
\usepackage{multirow}
\usepackage{bm}

\usepackage{float}
\usepackage{booktabs}
\usepackage{threeparttable}
\usepackage{ulem,soul}
\usepackage{xcolor}
\usepackage{arydshln}

\arxivnumber{} 
\title{\boldmath Dark energy perturbations and the robustness of cosmological 
neutrino-mass measurements}

\author[a, b]{Yu-Hang Yang,}
\author[c,b,d, 1]{Emmanuel N. Saridakis,\note{Corresponding author.}}
\author[a, b, 1]{Yi-Fu Cai,}
\author[e]{Hao-Ran Yu}

\affiliation[a]{Department of Astronomy, School of Physical Sciences, University of Science and Technology of China, Hefei 230026, China}

\affiliation[b]{CAS Key Laboratory for Research in Galaxies and Cosmology, School of Astronomy and Space Science, University of Science and Technology of China, Hefei 230026, China}

\affiliation[c]{National Observatory of Athens, Lofos Nymfon 11852, Greece}

\affiliation[d]{Departamento de Matem\'{a}ticas, Universidad Cat\'{o}lica del Norte, Avda. Angamos 0610, Casilla 1280, Antofagasta, Chile}

\affiliation[e]{Department of Astronomy, Xiamen University, Xiamen, Fujian 361005, China}

\emailAdd{yyh1024@mail.ustc.edu.cn}
\emailAdd{msaridak@noa.gr}
\emailAdd{yifucai@ustc.edu.cn}
\emailAdd{haoran@xmu.edu.cn}

\abstract{Cosmological observations are placing increasingly stringent bounds on the sum of neutrino masses, approaching the lower limits implied by neutrino oscillation experiments. Recent studies have suggested that dynamical dark energy may alleviate this apparent tension. However, these conclusions generally rely on the assumption that dark energy remains smooth, neglecting its perturbations.
In this work we investigate the robustness of cosmological neutrino-mass constraints by consistently incorporating dark-energy perturbations. Using CMB, BAO, RSD, and supernova data, we show that the commonly reported alleviation of the neutrino-mass tension in dynamical dark-energy models is not generic. While smooth dark energy substantially relaxes the neutrino-mass bounds, allowing dark energy to cluster shifts the preferred neutrino mass toward smaller, and even more negative, effective values.
We demonstrate that this behavior originates from a degeneracy between neutrino free-streaming and dark-energy perturbations in structure-growth observables. Different combinations of neutrino mass and dark-energy clustering can provide similarly good fits to current data while yielding significantly different neutrino-mass constraints. Our results show that cosmological neutrino-mass measurements are inherently model dependent and that reliable neutrino-mass inference requires a consistent treatment of dark-energy perturbations.}

\begin{document}
\maketitle
\flushbottom

\section{Introduction}

Cosmological observations are reaching a level of precision at which the
inferred bounds on the sum of neutrino masses approach, and in some cases
challenge, the lower limits implied by neutrino oscillation experiments. This
emerging tension has become one of the most intriguing issues at the interface
between particle physics and cosmology. However, its interpretation relies on
underlying assumptions about the dark sector, which may not be fully justified.

Neutrino oscillation experiments have established that neutrinos have mass
\cite{Capozzi:2021fjo,Esteban:2024eli,deSalas:2020pgw,PhysRevD.110.030001}.
Nevertheless, such experiments are only sensitive to the mass squared differences
between three different mass eigenstates, and thus cannot determine the
absolute neutrino mass scale. Constraining the absolute neutrino mass remains a
central challenge in modern physics, pursued through two complementary
avenues: laboratory-based experiments
\cite{KATRIN:2019yun,KATRIN:2024cdt} and cosmological observations
\cite{PhysRev.92.1347,Lesgourgues:2006nd}. Recently, a potential tension has
emerged between the neutrino mass constraints inferred from cosmological data
and those suggested by oscillation measurements.

Oscillation experiments \cite{PhysRevD.110.030001} have demonstrated that at
least two mass eigenstates are massive, and have precisely measured the squared
mass splittings, defined as $\Delta m_{ji}^2 \equiv m_j^2 - m_i^2$. The results
are $\Delta m_{21}^2 \approx 7.5 \times 10^{-5}~\mathrm{eV}^2$ and
$\Delta m_{32}^2 \approx 2.5 \times 10^{-3}~\mathrm{eV}^2$. These values imply
that $m_1$ and $m_2$ are nearly degenerate, while $m_3$ is either significantly
heavier (normal ordering, NO) or substantially lighter (inverted ordering, IO)
than the other two. As a result, oscillation data impose a lower bound on the
sum of neutrino masses: $\sum m_\nu > 0.05898~\mathrm{eV}$ for NO and
$\sum m_\nu > 0.09982~\mathrm{eV}$ for IO. In contrast, cosmological
observations provide upper limits on $\sum m_\nu$ by assuming minimal physical
priors (e.g., $\sum m_\nu > 0$). For instance, a recent combination of baryon
acoustic oscillation (BAO) measurements from the Dark Energy Spectroscopic
Instrument (DESI) Data Release 2, together with cosmic microwave background
(CMB) data from \textit{Planck} and the Atacama Cosmology Telescope (ACT),
yields the strongest current astrophysical upper limit of
$\sum m_\nu < 0.0642$ eV (95\% C.L.), under the assumption of the cosmological
$\Lambda$CDM model and three degenerate neutrino states
\cite{Elbers:2025vlz}. This constraint is intriguingly close to, and in the
case of IO, even below, the lower bounds required by oscillation experiments,
highlighting a potential tension between cosmological and terrestrial probes of
neutrino masses, see
\cite{Craig:2024tky,Lynch:2025ine,Jiang:2024viw,Giare:2025ath,Gorbunov:2026sly,
Montandon:2026vuc,Yang:2026yaq,Novell-Masot:2026sgg,Pulido-Hernandez:2026hcs,
Feng:2026pzs,Du:2025xes,Li:2026asg} for more discussions.

Meanwhile, recent results from DESI DR2 \cite{DESI:2025zgx} favor a dynamical
dark-energy model when interpreted within the $w_0w_a$
(Chevallier-Polarski-Linder) parameterization
\cite{Chevallier:2000qy,Linder:2002et}, where the equation of state is given by
$w(a)=w_0+w_a(1-a)$. Depending on the choice of supernova sample, the
significance of the deviation from the cosmological constant ranges from
$2.8$ to $4.2\sigma$, hinting that the origin of cosmic acceleration may
extend beyond the simple cosmological constant. A natural interpretation of
this situation invokes dynamical dark energy, which is known to be degenerate
with neutrino mass in cosmological observables. Intriguingly, a time-varying
dark-energy component within the $w_0w_a$CDM model has been argued to
alleviate the preference for a negative effective neutrino mass, due to the
degeneracy between neutrino mass and the background evolution of dark energy
\cite{Hannestad:2005gj,Dirian:2017pwp,RoyChoudhury:2019hls,
Upadhye:2017hdl,Vagnozzi:2018jhn,RoyChoudhury:2024wri,Chebat:2025kes}.
However, most existing analyses focus primarily on the background expansion
history. Since cosmological neutrino-mass constraints are driven predominantly
by large-scale-structure observables, a consistent assessment of their
robustness requires examining not only the background evolution of dark energy
but also its perturbations.

Nevertheless, this conclusion hinges on a crucial and often overlooked assumption:
that dark energy behaves as a smooth component on sub-horizon scales, i.e.,
that its perturbations can be neglected. This assumption is nontrivial. In
general, dark energy may cluster depending on its effective sound speed and
equation of state, and its perturbations can significantly affect the growth of
cosmic structures
\cite{Takada:2006xs,Basse:2012wd,Batista:2021uhb,Kunz:2015oqa}. In
particular, dark-energy perturbations can suppress the growth of structure on
small scales, an effect that can partially mimic or compensate for the
suppression induced by massive neutrinos
\cite{Takada:2006xs,Takada:2005si}. As a result, an additional degeneracy
arises at the perturbative level, which is not captured when dark energy is
treated as smooth. Consequently, cosmological constraints on neutrino mass may
depend sensitively on the assumed perturbative properties of dark energy.
Despite the extensive literature on neutrino masses in dynamical dark-energy
cosmologies, the impact of dark-energy perturbations on the robustness of
cosmological neutrino-mass measurements has not been systematically quantified.
This leaves open the possibility that part of the apparent preference for
specific neutrino-mass values originates from assumptions regarding the
perturbative behavior of the dark-energy sector rather than from the neutrino
sector itself.

In this work, we critically reassess the robustness of cosmological
neutrino-mass measurements by consistently incorporating dark-energy
perturbations. We investigate the interplay between clustering dark energy and
massive neutrinos, and assess whether the apparent alleviation of the
neutrino-mass tension in dynamical dark-energy models persists once the full
perturbative dynamics are taken into account. We find that the commonly
reported alleviation of the neutrino-mass tension in dynamical dark-energy
models is not a robust prediction. Once dark-energy perturbations are
consistently included, the inferred neutrino-mass constraints can shift
significantly and the preference for smaller effective neutrino masses can be
partially restored. This behavior originates from a physical degeneracy between
neutrino free-streaming and dark-energy perturbations in structure-growth
observables. These results highlight that cosmological constraints on neutrino
mass are inherently model dependent, and that a consistent treatment of
dark-energy perturbations is essential for reliable neutrino-mass inference.

The structure of this paper is the following. In Sec.~\ref{framework} we
present the physical framework describing massive neutrinos and clustering dark
energy, emphasizing the degeneracies that arise at the perturbation level. In
Sec.~\ref{dataresults} we describe the methodology and data sets used in our
analysis, and we present the resulting constraints on the neutrino mass,
highlighting the impact of dark-energy perturbations. Finally, in
Sec.~\ref{sec:concl} we summarize our findings and discuss their implications
for the robustness of cosmological neutrino-mass measurements.

\section{Physical framework: massive neutrinos and clustering dark energy}
\label{framework}

In this section we present the theoretical framework describing the effects of 
massive neutrinos and dark energy perturbations on cosmological evolution. In 
particular, we emphasize their distinct physical mechanisms and the 
degeneracies that arise at the perturbative level, which play a central role in 
the interpretation of cosmological neutrino-mass measurements.

\subsection{Massive neutrinos in cosmology}

Massive neutrinos influence cosmic evolution in two distinct ways. First, they affect the expansion history of the Universe. In the early Universe, neutrinos are highly relativistic and behave as radiation. As the Universe expands and cools, their thermal momenta redshift away. When the average momentum drops below the neutrino mass, satisfying $\langle p_{\nu} \rangle = 3T_{\nu} \simeq m_{\nu}$, neutrinos transition to a non-relativistic state. At late times, they effectively behave as a dark matter component, contributing to the total matter density. This effect can be probed through background observables such as baryon acoustic oscillations (BAO).

Second, massive neutrinos suppress density perturbations on small scales, with $\Delta P_\mathrm{m} / P_\mathrm{m} \approx -8 f_\nu$, where $f_\nu \equiv \Omega_\nu / \Omega_\mathrm{m}$ is the neutrino mass fraction. This suppression arises because the large thermal velocities of neutrinos prevent them from clustering below their characteristic free-streaming length. On scales smaller than the free-streaming scale, neutrinos cannot be confined within collapsing overdensities, leading to a damping of structure formation. This results in a characteristic scale-dependent suppression in the matter power spectrum. Redshift-space distortion (RSD) measurements are particularly sensitive to this effect, as they directly probe the growth rate of cosmic structures.

Traditionally, cosmic neutrinos are modeled as massive particles that play a 
dual role in cosmological evolution. The energy density of massive neutrinos is 
given by
\begin{equation}
    \rho_{\nu} = \sum_{i}^{N_{\nu}} \frac{g_i(1+z)^4}{2\pi^2} 
\int_{0}^{\infty}\frac{p^2 \epsilon(p,m_i)}{1+e^{p/T_{\nu,0}}} \mathrm{d}p,
\end{equation}
where $N_{\nu}$ is the number of neutrino species, $g_i$ is the degeneracy 
factor and $m_i$ the mass of neutrino species $i$. The neutrino energy is $\epsilon(p, m) = \sqrt{p^2 + a^2 m^2}$, and the present-day neutrino temperature is $T_{\nu,0}$. Consequently, massive neutrinos contribute to the current energy density as $\Omega_{\nu} \simeq \sum m_{\nu} / 93.14 \, h^2$.

However, current cosmological data exhibit a preference for a very small, or even negative, sum of neutrino masses. This unphysical preference points to a ``prior weight effect'', wherein the physical prior $\sum m_\nu > 0$ biases the posterior when it forces the latter away from its maximum likelihood value. To circumvent this issue, recent works have allowed $\sum m_\nu$ to take negative values in Bayesian analyses, thereby defining an effective neutrino mass \cite{Craig:2024tky, Green:2024xbb, Elbers:2024sha}. Note that the negative mass is merely a diagnostic variable revealing a tension between data and model assumptions. In this framework, the neutrino energy is replaced by
\begin{equation}
    \epsilon_{\mathrm{eff}} = \mathrm{sgn}(m_{\nu,\mathrm{eff}}) \sqrt{p^2 + a^2 m_{\nu,\mathrm{eff}}^2},
\end{equation}
such that negative values effectively enhance small-scale perturbations, in contrast to the suppression induced by positive masses.

\subsection{Dark energy perturbations}

Similar to massive neutrinos, dark energy can affect cosmological evolution 
both at the background and perturbation levels. In general, dark energy can be 
modeled as an effective fluid whose perturbations are non-adiabatic and may 
carry entropy (isocurvature) fluctuations. In an arbitrary gauge, the relation 
between pressure and density perturbations is given by
\begin{equation}
\delta p_{\rm de}=c_s^2\delta\rho_{\rm 
de}+3aH(1+w)\left(c_s^2-c_a^2\right)\rho_{\rm de}\frac{\theta_{\rm de}}{k^2},
\end{equation}
where $c_s^2$ is the effective sound speed in the dark energy rest frame, and 
$c_a^2=\dot p_{\rm de}/\dot\rho_{\rm de}$ is the adiabatic sound speed.

The clustering behavior of dark energy is governed by its sound speed through 
the sound horizon
\begin{equation}
    r_s = \int_{0}^a \frac{c_s}{a^2H}\mathrm{d} a \sim \frac{c_s}{aH}.
\end{equation}
Below this scale, pressure support suppresses the growth of dark energy 
perturbations. For instance, $c_s = 10^{-3}$ corresponds to $r_s \simeq 
5~\mathrm{Mpc}$, while $c_s=10^{-5}$ yields $r_s \simeq 0.05~\mathrm{Mpc}$.

In the matter-dominated era, and in the clustering limit $c_s^2 \ll 1$, dark 
energy perturbations for constant $w$ are related to matter perturbations as 
\cite{Abramo:2008ip,Sapone:2009mb,Ballesteros:2010ks,Batista:2021uhb}
\begin{equation}
    \delta_{\mathrm{de}} = \frac{1+w}{1-3w}\delta_{\mathrm{m}},
    \label{eq:delta_de}
\end{equation}
as we show in   Appendix~\ref{app:de_per}. This relation shows that the impact of dark energy perturbations depends on 
both $w$ and $c_s$. 
In principle, the dark-energy sound speed may be treated as a free parameter.
However, current cosmological observations provide only weak constraints on
$c_s^2$, and its correlations with the other cosmological parameters remain
limited. For completeness, we present the corresponding analysis in Appendix \ref{app:free cs2}. For $w \simeq -1$, dark energy perturbations are strongly 
suppressed regardless of $c_s$. For $c_s^2\simeq1$, dark energy remains 
effectively smooth on sub-horizon scales, while for $c_s^2\ll1$ it can cluster 
and affect the growth of matter perturbations and gravitational potentials.

Consequently, observables such as CMB anisotropies, the matter power spectrum, 
and the growth rate $f\sigma_8$ can be significantly modified by the presence 
of clustering dark energy 
\cite{Batista:2014uoa,Mehrabi:2015hva,Mehrabi:2014ema}. In particular, for $w > 
-1$ dark energy perturbations enhance structure formation, while for $w < -1$ 
they suppress it.

Early analyses, such as those by the \textit{Planck} collaboration 
\cite{Planck:2015bue}, found the sound speed $c_s$ to be largely unconstrained, 
mainly because they assumed a constant equation of state with $w \approx -1$, 
where perturbations are negligible. However, recent DESI results 
\cite{DESI:2025zgx,Yang:2025gaz} favor a time-evolving dark energy, opening the possibility 
of probing its clustering properties observationally.

Another alternative is to describe dark energy within the Effective Field 
Theory of dark energy (EFTofDE) framework 
\cite{Gubitosi:2012hu,Bloomfield:2012ff,Creminelli:2008wc,Gleyzes:2013ooa}, 
where perturbations are parametrized through time-dependent functions 
$\alpha_i$. In this approach,
one works in unitary gauge and writes the most general action compatible with
the residual spatial diffeomorphism invariance. The resulting action is
organized in terms of operators constructed from the metric, the extrinsic
curvature, and their perturbations, whose coefficients are arbitrary functions
of time. This provides a unified description of a broad class of single-field
dark-energy and modified-gravity models. For Horndeski-like theories, the information contained in these EFT coefficients
at the level of linear cosmological perturbations can be reorganized into a
more physically transparent basis \cite{Bellini:2014fua,Hu:2014oga}, called $\alpha$-basis. In this $\alpha$-basis, the 
relevant functions are $\alpha_{\rm M}$, $\alpha_{\rm B}$, $\alpha_{\rm K}$, 
and $\alpha_{\rm T}$:
\begin{equation}
\begin{aligned}
\alpha_{\rm M} &\equiv \frac{\mathrm{d}\ln M_*^2}{\mathrm{d}\ln a}, \qquad
\alpha_{\rm B} \equiv -\frac{M_{\rm P}^2\dot\Omega+\bar M_1^3}{H M_*^2}, \\
\alpha_{\rm K} &\equiv \frac{2c+4M_2^4}{H^2 M_*^2}, \qquad
\alpha_{\rm T} \equiv c_T^2-1 = \frac{M_3^2}{M_*^2},
\end{aligned}
\end{equation}
where $M_*^2(t)=M_{\rm P}^2[1+\Omega(t)]-\bar M_3^2$ is the effective Planck 
mass, $\Omega$, $M_1$, $M_2$, $M_3$ are functions of time, $c$ is the speed of light and $c_T$ is the tensor propagation speed. The four functions respectively characterize
the running of the effective Planck mass, the kinetic mixing (``braiding'')
between the scalar and metric degrees of freedom, the scalar kinetic
contribution (``kineticity''), and the tensor propagation speed. This
$\alpha$-basis is also used in \texttt{hi\_class}, which we employ for the
numerical evolution of the linear perturbations
\cite{Zumalacarregui:2016pph,Bellini:2019syt}.

In general, the $\alpha_i$ are arbitrary functions of time. In order to obtain
a tractable phenomenological description for parameter inference, we adopt
the commonly used parametrization
\begin{equation}
    \alpha_i(a)=c_i\Omega_\mathrm{de}(a),
    \label{eq:alpha_parameterization}
\end{equation}
with constant coefficients $c_i$ \cite{Pujolas:2011he, Barreira:2014jha, 
Bellini:2015oua}. This prescription has the useful property
that modifications to gravity are naturally suppressed at early times, when
$\Omega_{\rm de}(a)\ll1$, while they become relevant mainly at late times,
when dark energy contributes significantly to the cosmic energy budget.
Thus, the standard early-Universe evolution is approximately recovered,
while the late-time perturbative freedom relevant for structure formation is
retained. We emphasize that Eq.~\eqref{eq:alpha_parameterization} is a
phenomenological parametrization rather than a unique fundamental model; here
it is employed as a robustness test of the inferred neutrino-mass constraints
against additional perturbative degrees of freedom in the dark sector. 

We set $\alpha_T=0$, motivated by the stringent constraints on the
propagation speed of gravitational waves \cite{LIGOScientific:2017zic}, and
fix $c_K=10^{-2}$ since current cosmological observations are only weakly
sensitive to the kineticity parameter \cite{Bellini:2015xja,Reischke:2018ooh}.
Stability conditions are imposed throughout the parameter exploration to
avoid ghost and gradient instabilities, while the background expansion is
taken to follow the $w_0w_a$CDM model.

\subsection{Degeneracy between neutrinos and dark energy perturbations}
\label{subsectiondegeneracy}

Both massive neutrinos and dark energy perturbations affect the formation and
evolution of cosmic structures, although through distinct physical mechanisms.
Massive neutrinos suppress the growth of matter perturbations below their
free-streaming scale, while dark energy perturbations modify the evolution of
gravitational potentials and the growth rate of structures. Depending on the
equation of state and clustering properties of dark energy, these effects can
either enhance or suppress structure formation.

The key point is that cosmological observations do not directly measure either
the neutrino mass or the dark-energy perturbations themselves. Instead, they
probe quantities such as the matter power spectrum, the growth rate
$f\sigma_8$, and weak-lensing observables, all of which depend on the overall
evolution of matter perturbations and gravitational potentials. Consequently,
different physical mechanisms that modify the growth history in a similar
manner can lead to nearly indistinguishable observational signatures.

For massive neutrinos, the suppression of growth originates from their large
thermal velocities. Below the free-streaming scale, neutrinos do not cluster
efficiently and therefore contribute less to the gravitational potentials that
drive structure formation. The resulting reduction in the growth of matter
perturbations constitutes one of the primary signatures through which
cosmological observations constrain the neutrino mass. On the other hand,
clustering dark energy modifies the evolution of the gravitational potentials
through its own density perturbations. In the clustering regime
($c_s^2\ll 1$), dark energy participates in the growth process and can
significantly alter the evolution of matter perturbations. Depending on the
equation of state, this modification can either enhance or suppress the growth
of structure.

As a result, a nontrivial degeneracy arises at the perturbation level. In
particular, clustering dark energy with $w<-1$ tends to suppress the growth of
structures, producing signatures that can partially mimic those of massive
neutrinos. More generally, the effects of neutrino free-streaming and dark
energy clustering can compensate each other, leading to similar predictions for
observables sensitive to structure growth. Consequently, measurements of the
matter power spectrum and the growth rate $f\sigma_8$ cannot, in general,
unambiguously disentangle these contributions without a consistent treatment of
both sectors.

This degeneracy is particularly relevant for the growth information entering
cosmological neutrino-mass constraints. In practice, the data determine a preferred growth
history, and several combinations of neutrino masses and dark-energy
perturbation properties can reproduce this history with comparable accuracy.
The inferred neutrino mass therefore does not depend solely on the neutrino
sector, but also on the assumptions adopted for the perturbative behavior of
dark energy. From this perspective, analyses that neglect dark-energy
perturbations effectively remove part of the physically allowed parameter space
by assumption, thereby breaking the degeneracy a priori and potentially
biasing the resulting neutrino-mass constraints. We note that the cosmological effects of neutrino masses can be
phenomenologically separated into background and perturbative contributions,
as discussed in \cite{Bertolez-Martinez:2024wez}. In such a description,
the background contribution controls the homogeneous neutrino energy density,
while the perturbative contribution governs the free-streaming properties of
neutrino perturbations. Since the degeneracy studied here arises from the
interplay between dark-energy clustering and neutrino free-streaming in the
growth sector, we expect our qualitative conclusion to remain valid under
such a separation. Nevertheless, the corresponding quantitative constraints
on the two effective neutrino-mass contributions may differ and would require
a dedicated analysis.

\begin{figure}[ht!]
    \centering
    \includegraphics[width=0.6\textwidth]{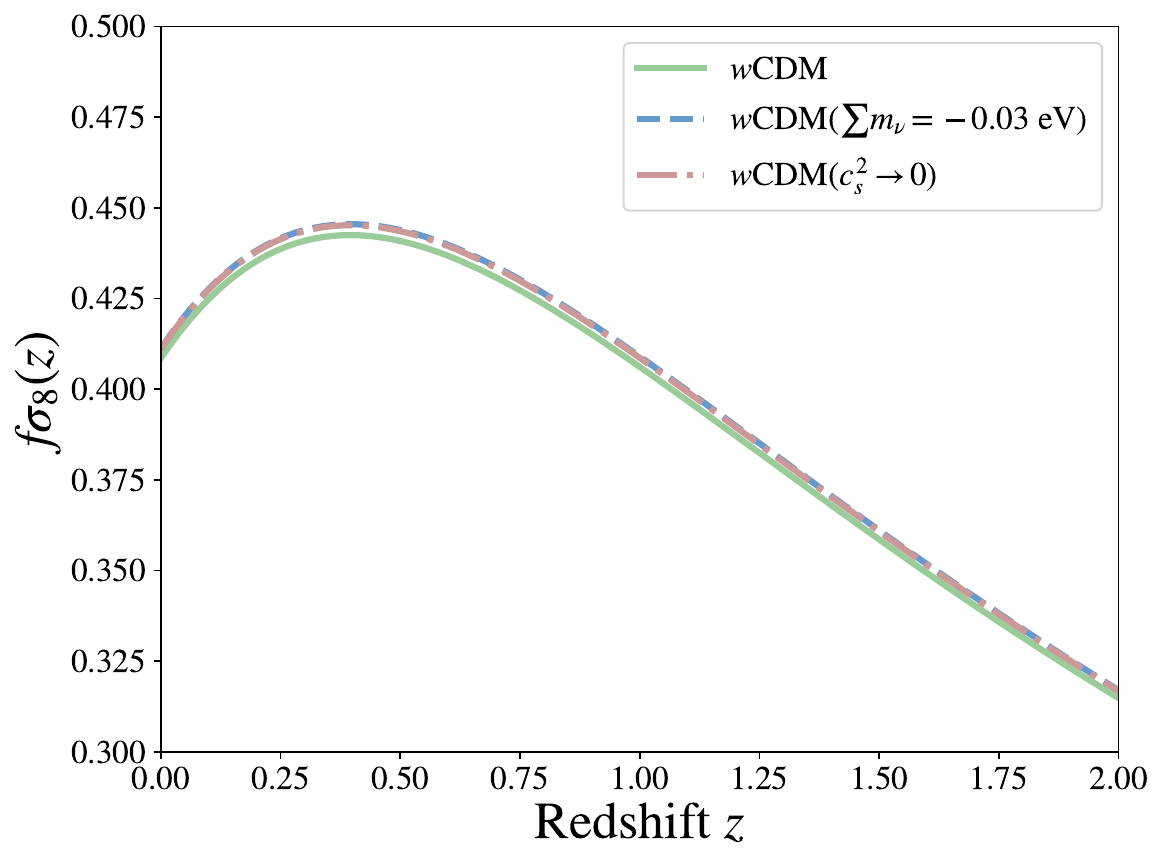}    
    \caption{{\it{A schematic illustration of the degeneracy between neutrino free-streaming and dark-energy perturbations. Different combinations of effective neutrino mass and dark-energy clustering can produce similar modifications to the growth of cosmic structures. The example shown corresponds to $w=-0.85$.}}}
    \label{fig:schematic_wcdm_growth}
\end{figure}

This observation has direct implications for recent claims that dynamical
dark-energy models alleviate the tension between cosmological neutrino-mass
measurements and the lower bounds implied by oscillation experiments. If the
apparent preference for larger neutrino masses originates partly from the
assumption of smooth dark energy, then the inferred alleviation may not be a
robust prediction of the underlying cosmological model. Establishing whether
this is indeed the case constitutes the main objective of the present work.

A schematic illustration of this interplay is presented in
Fig.~\ref{fig:schematic_wcdm_growth}. As shown, clustering dark energy and the
effective neutrino mass can induce qualitatively similar modifications to the
growth history, allowing one effect to partially compensate for the other in
cosmological observables. This physical degeneracy forms the basis of the
analysis presented in the following sections. Its impact on cosmological
parameter estimation will be quantified in Sec.~\ref{dataresults}, while its
origin in the underlying observables will be illustrated explicitly in
Appendix~\ref{app:separate_effects}.

\section{Methodology, data, and results}
\label{dataresults}

In this section, we present the methodology and data employed in our analysis, 
and we derive the resulting constraints on the neutrino mass. In particular, we 
investigate how the inclusion of dark energy perturbations affects the inferred 
values, thereby assessing the robustness of cosmological neutrino-mass 
measurements. Within the framework of massive neutrinos or dark energy perturbations, the growth rate $f(z)$ will no longer be scale-independent. The RSD data should be understood as the measurements which are scale-averaged over the observed $k$-range.

\subsection{Methodology and data}

We perform a Monte Carlo Markov Chain (MCMC) analysis to constrain the 
effective neutrino mass. The background and linear perturbation evolution of 
the cosmological models are computed using modified versions of \texttt{hi\_class}~\cite{Bellini:2019syt,Zumalacarregui:2016pph,Blas2011TheCL}. Furthermore, parameter 
estimation is carried out with the public sampler 
\texttt{MontePython}~\cite{Audren:2012wb,Brinckmann:2018cvx}, and convergence is 
assessed using the Gelman-Rubin diagnostic, requiring $R-1<0.02$ for all 
chains.

Our data combination includes:
\begin{itemize}
\item The full \textit{Planck} CMB temperature and polarization spectra 
\cite{Planck:2018vyg}, together with lensing measurements from both ACT DR6 
\cite{AtacamaCosmologyTelescope:2025blo} and \textit{Planck} DR4 
\cite{Carron:2022eyg}. For high-$\ell$ TTTEEE likelihoods we use the 
\texttt{CamSpec} likelihood \cite{Efstathiou:2019mdh,Rosenberg:2022sdy}, and 
for low-$\ell$ temperature and polarization we use \texttt{Commander} and 
\texttt{SimAll}.
\item The DES-Dovekie dataset, based on a comprehensive recalibration and 
reanalysis of the DESY5 supernova sample \cite{DES:2025sig}.
\item Baryon acoustic oscillation (BAO) and redshift-space distortion (RSD) 
measurements from Sloan Digital Sky Survey’s (SDSS)
Baryon Oscillation Spectroscopic Survey (BOSS) and Extended
Baryon Oscillation Spectroscopic Survey (eBOSS), including SDSS DR7 main
galaxy sample \cite{Ross:2014qpa,Howlett:2014opa}, BOSS 
DR12 \cite{BOSS:2016wmc}, and eBOSS DR16 data for ELGs \cite{eBOSS:2020qek,eBOSS:2020abk,eBOSS:2020fvk}, LRGs \cite{eBOSS:2020lta,eBOSS:2020hur}, QSOs \cite{eBOSS:2020uxp,eBOSS:2020gbb}, and the 
Lyman-$\alpha$ forest \cite{eBOSS:2020fvk}.
\end{itemize}

\begin{table}[ht]
\centering
\small
\caption{Cosmological and nuisance parameters varied in the analysis, together with their prior ranges. All priors are uniform over the intervals shown.}
\begin{tabular}{cccc}
\hline
\textbf{Model} &\textbf{Parameter} & \textbf{Default} & \textbf{Prior}\\
\hline

Base &$\omega_\mathrm{b}$ & - & $\mathcal{U}[0.005, 0.1]$ \\
&$\omega_\mathrm{cdm}$ & - & $\mathcal{U}[0.001, 0.99]$  \\
&$100\theta_\mathrm{MC}$ & - & $\mathcal{U}[0.5, 10]$  \\
&$\ln(10^{10}A_s)$ & - & $\mathcal{U}[1.61, 3.91]$  \\
&$n_s$ & - & $\mathcal{U}[0.8, 1.2]$  \\
&$\tau$ & - & $\mathcal{U}[0.01, 0.8]$  \\
&$\sum m_{\nu}$ & 0.06 & $\mathcal{U}[-5,5]$ \\
\hline
DE&$w_0$ or $w$ & $-1$ & $\mathcal{U}[-3, 1]$ \\
&$w_a$ & 0 & $\mathcal{U}[-3, 2]$ \\
\hline
EFT&$c_\mathrm{B}$ & 0 & $\mathcal{U}[-10, 10]$  \\
&$c_\mathrm{M}$ & 0 & $\mathcal{U}[-10, 10]$  \\
\hline
\end{tabular}
\label{table:mcmc_para}
\end{table}

Our baseline analysis adopts the approximation
$\sum m_\nu=3m_\nu$, corresponding to three degenerate neutrino mass
eigenstates. This choice is motivated by the limited sensitivity of current
cosmological observations to the neutrino mass hierarchy
\cite{Vagnozzi:2017ovm,Archidiacono:2020dvx}. Throughout the analysis, we assume the Standard Model prediction
$N_{\rm eff}=3.044$ \cite{Froustey:2020mcq,Bennett:2020zkv,Drewes:2024wbw}. The effective neutrino energy is defined as
$
    \epsilon_{\rm eff}
    =
    {\rm sgn}(m_{\nu,\rm eff})
    \sqrt{p^2+a^2m_{\nu,\rm eff}^2},
$
so that the contribution of the effective massive-neutrino sector to the
energy-momentum tensor changes sign when $m_{\nu,\rm eff}<0$. As a
consequence, its relativistic contribution at early times also changes sign. In order to keep the total relativistic energy density, and hence the
early-Universe expansion history, fixed when crossing
$\sum m_{\nu,\rm eff}=0$, the number of massless neutrino species
$N_{\rm ur}$ is adjusted accordingly. With the \texttt{CLASS} convention,
we use $N_\mathrm{ur}=0.00441$ for $\sum m_\nu>0$ and $N_\mathrm{ur}=6.08627$ for $\sum m_\nu<0$, which preserves $N_{\rm eff}=3.044$ in both branches \cite{Elbers:2025vlz}. We stress that the
value $N_{\rm ur}=6.08627$ does not represent six additional physical
neutrino species. It is a bookkeeping prescription required by the
phenomenological continuation to negative effective neutrino masses, ensuring
that the early-time radiation density remains unchanged.

The full set of cosmological parameters and their corresponding prior ranges is summarized in Table~\ref{table:mcmc_para}. We consider two background cosmologies, namely $\Lambda$CDM and $w_0w_a$CDM. The results for the $w$CDM model are not included in the main text,
since the proximity of the preferred values of $w$ to $-1$
renders the impact of dark-energy clustering on the inferred
neutrino mass negligible.  For completeness, the corresponding results are presented in Appendix~\ref{app:wcdm}.

\subsection{Results}

We now present the constraints on the effective neutrino mass and examine how
they are modified when dark-energy perturbations are consistently taken into
account. Our primary goal is to assess the robustness of the widely discussed
claim that dynamical dark-energy models alleviate the tension between
cosmological neutrino-mass measurements and the lower bounds implied by
oscillation experiments.
 
As we will show, the inferred neutrino-mass constraints depend not only on the
background dark-energy evolution, but also on the perturbative properties of
the dark-energy sector. In particular, the inclusion of dark-energy clustering
introduces a nontrivial degeneracy with neutrino free-streaming, which
substantially alters the preferred neutrino-mass range.

\begin{figure}[htbp]
    \centering
    \begin{minipage}{0.49\textwidth}
        \centering
        \includegraphics[width=\textwidth]{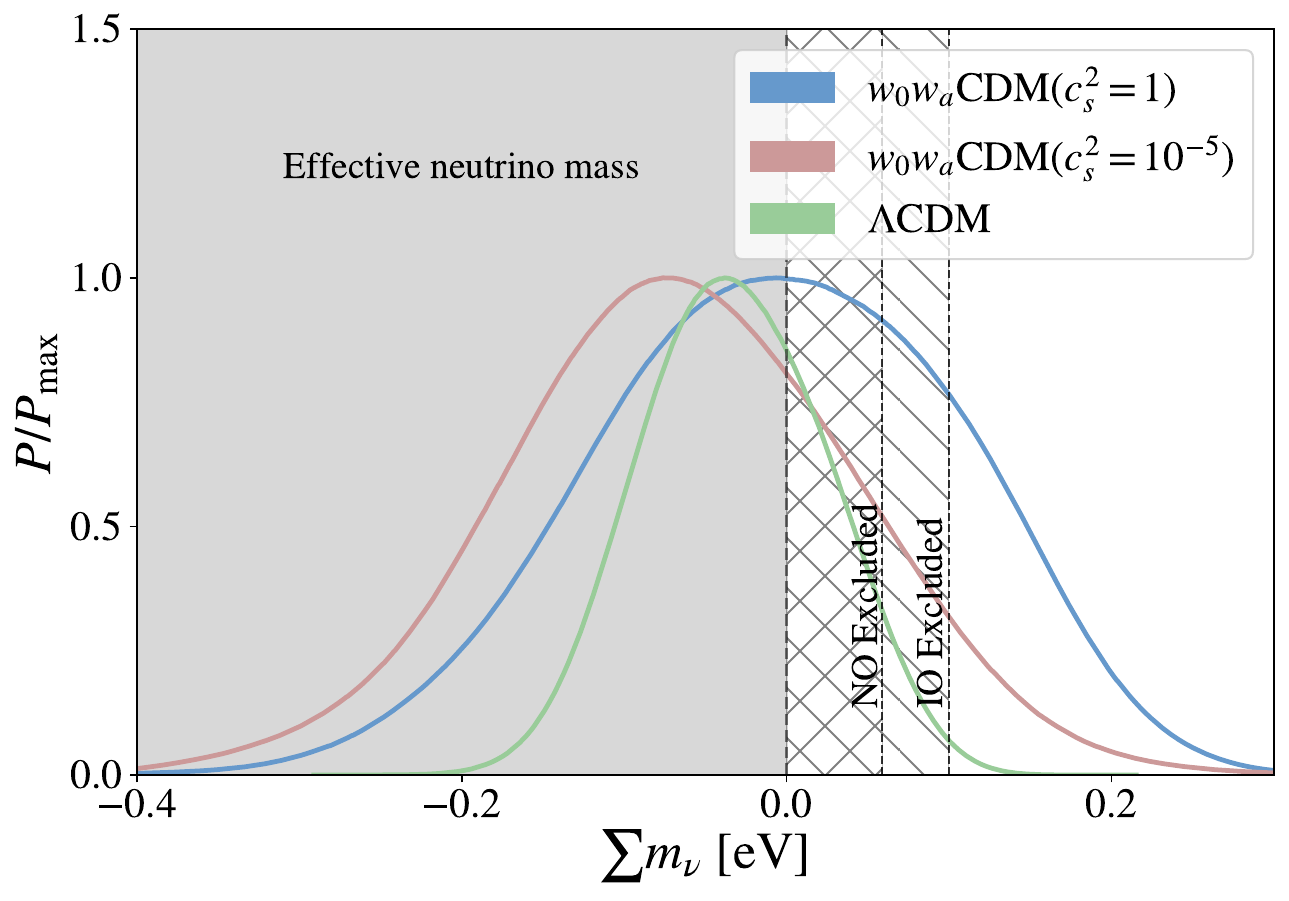}
    \end{minipage}
    \hfill
    \begin{minipage}{0.49\textwidth}
        \centering
        \includegraphics[width=\textwidth]{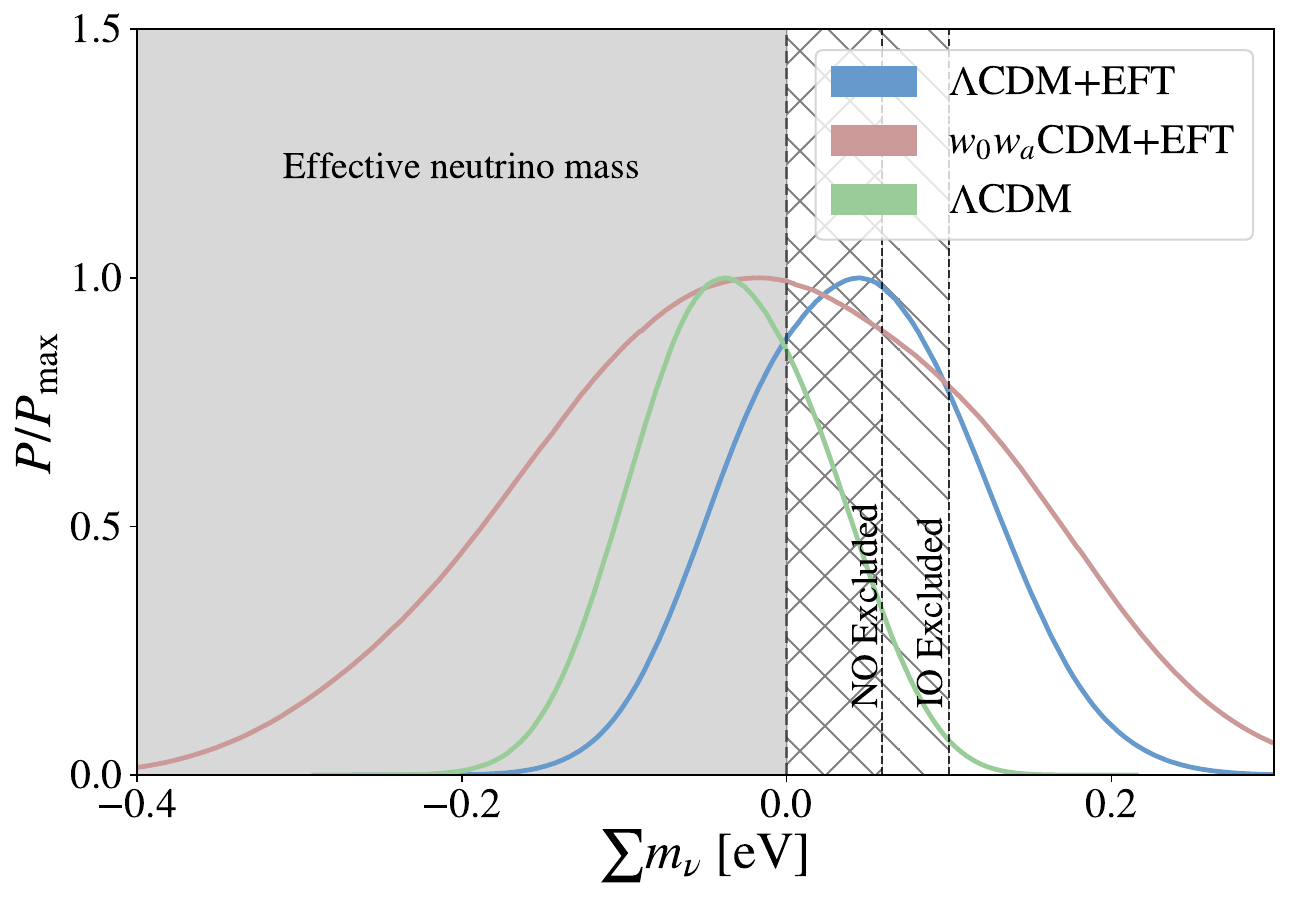}
    \end{minipage}   
    \caption{{\it{Normalized one-dimensional posterior distributions of the effective neutrino mass
$\sum m_\nu$ for the $\Lambda$CDM and $w_0w_a$CDM models, obtained from the
combined BAO+CMB+SNe data set. The lower bounds implied by neutrino oscillation
measurements are also shown for both the normal ordering (NO) and inverted
ordering (IO). The left panel presents the results for smooth and clustering
dark-energy models, while the right panel shows the corresponding constraints
within the effective field theory (EFT) framework.}}}
    \label{fig:nv_prob}
\end{figure}

Fig. \ref{fig:nv_prob} shows the marginalized posterior distribution of the
neutrino mass sum $\sum m_\nu$. For the $\Lambda$CDM model, the data prefer a
best-fit value corresponding to a negative effective neutrino mass. This
preference yields the upper limit
\begin{equation}
\Lambda \text{CDM}: \ \sum m_{\nu}<0.0634\ \text{eV (95\%)}.
\nonumber
\end{equation}
This limit lies below the lower bound implied by oscillation experiments for
the inverted ordering and is also very close to the lower bound for the normal
ordering, thereby highlighting the existing tension.

In the left panel of Fig. \ref{fig:nv_prob}, allowing for dynamical dark
energy within the $w_0w_a$CDM framework with smooth dark energy
($c_s^2=1$), relaxes the constraint to
\begin{equation}
w_0w_a \text{CDM}(c_s^2=1): \ \sum m_{\nu}<0.1696\ \text{eV (95\%)}.
\nonumber
\end{equation}
In this case, the upper limit is fully consistent with the lower bounds from
oscillation experiments for both mass orderings, indicating an apparent
alleviation of the tension. However, when dark-energy perturbations are
included and clustering is allowed ($c_s^2=10^{-5}$), the constraint becomes
\begin{equation}
w_0w_a \text{CDM}(c_s^2=10^{-5}): \ \sum m_{\nu}<0.1125\ \text{eV (95\%)}.
\nonumber
\end{equation}

The comparison between the smooth and clustering dark-energy cases reveals the
central result of this work. Although both models share the same background
parameterization, their inferred neutrino-mass constraints differ
significantly. The apparent alleviation of the neutrino-mass tension observed
in the smooth-dark-energy case is therefore not a generic consequence of a
dynamical dark-energy background. Instead, it depends crucially on the
assumption that dark energy remains unclustered.

Once dark-energy perturbations are included, the preferred neutrino mass shifts
back toward smaller values, partially restoring the preference already present
in $\Lambda$CDM. This demonstrates that neutrino-mass inference is sensitive to
the perturbative description of the dark-energy sector and that neglecting
dark-energy perturbations can lead to overly optimistic conclusions regarding
the resolution of the neutrino-mass tension.

The origin of this behavior can be traced directly to the perturbation-level
degeneracy discussed in subsection \ref{subsectiondegeneracy}. A negative effective neutrino mass enhances
the growth of matter perturbations, while clustering dark energy with
$w<-1$ suppresses structure formation. Since both effects act on the same
growth observables, a partial compensation becomes possible. Consequently, the
likelihood can be improved by shifting the preferred neutrino mass toward more
negative values when dark-energy clustering is present. The resulting posterior
shift is therefore not accidental, but reflects a genuine physical degeneracy
between neutrino free-streaming and dark-energy perturbations.

We proceed by providing the results for the EFTofDE. Unlike the effective fluid
description, the EFT framework allows for a broader range of perturbative
effects and can also encompass modified-gravity phenomena. Consequently, it
introduces additional degrees of freedom capable of absorbing part of the
observational preference associated with the effective neutrino mass.

Within the EFTofDE framework, the neutrino-mass tension is alleviated even for
the $\Lambda$CDM background:
\begin{equation}
\Lambda\text{CDM+EFT}: \ \sum m_{\nu}<0.1610\ \text{eV (95\%)},
\nonumber
\end{equation}
which is fully consistent with the bounds from oscillation experiments. This
alleviation arises because the additional EFT degrees of freedom can partially
absorb the observational preference for a negative effective neutrino mass.
Consequently, the inferred upper limit on $\sum m_\nu$ shifts upward, reducing
the tension with laboratory measurements. For the $w_0w_a$CDM+EFT case, the
corresponding constraint is
\begin{equation}
w_0w_a\text{CDM+EFT}: \ \sum m_{\nu}<0.2058\ \text{eV (95\%)},
\nonumber
\end{equation}
which is likewise fully compatible with the oscillation bounds.

\begin{figure}[htbp]
    \centering
    \centering
    \begin{minipage}{0.49\textwidth}
        \centering
        \includegraphics[width=\textwidth]{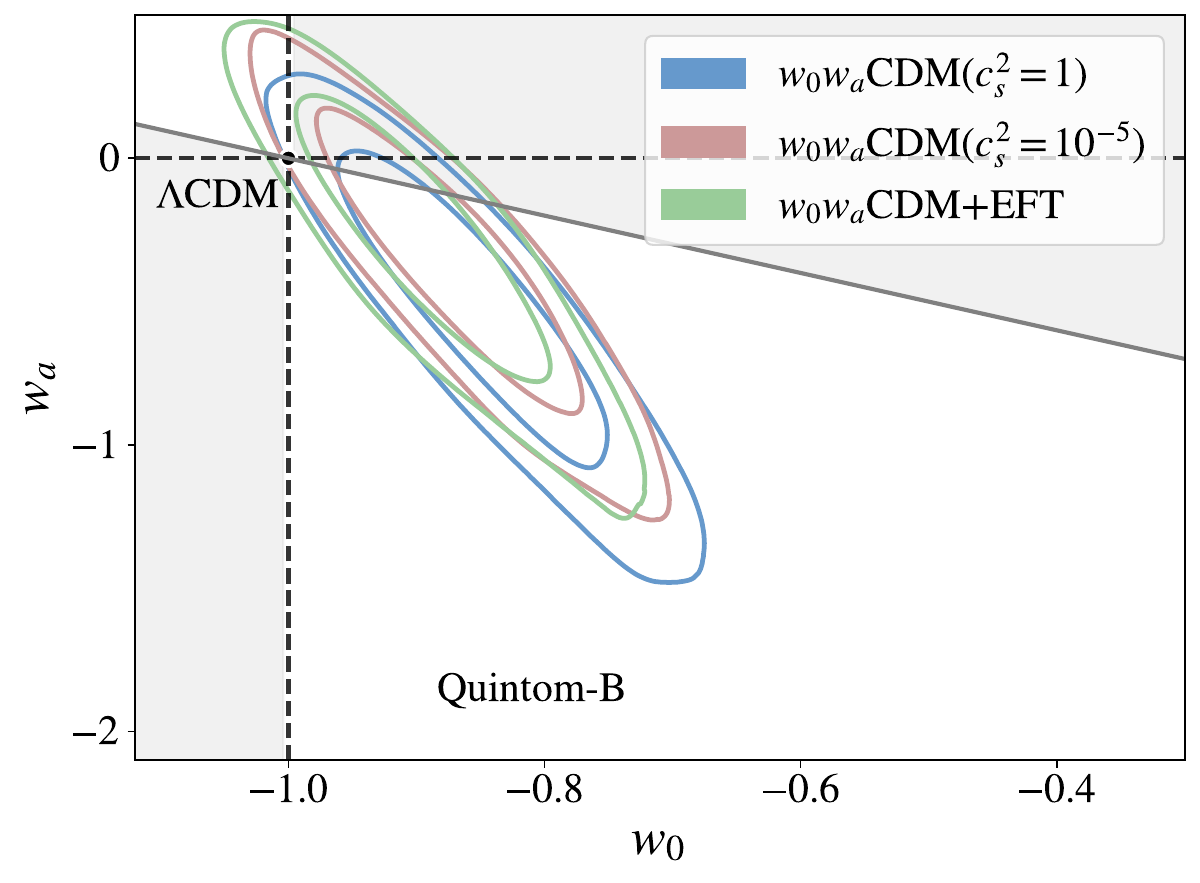}
    \end{minipage}
    \hfill
    \begin{minipage}{0.495\textwidth}
        \centering
        \includegraphics[width=\textwidth]{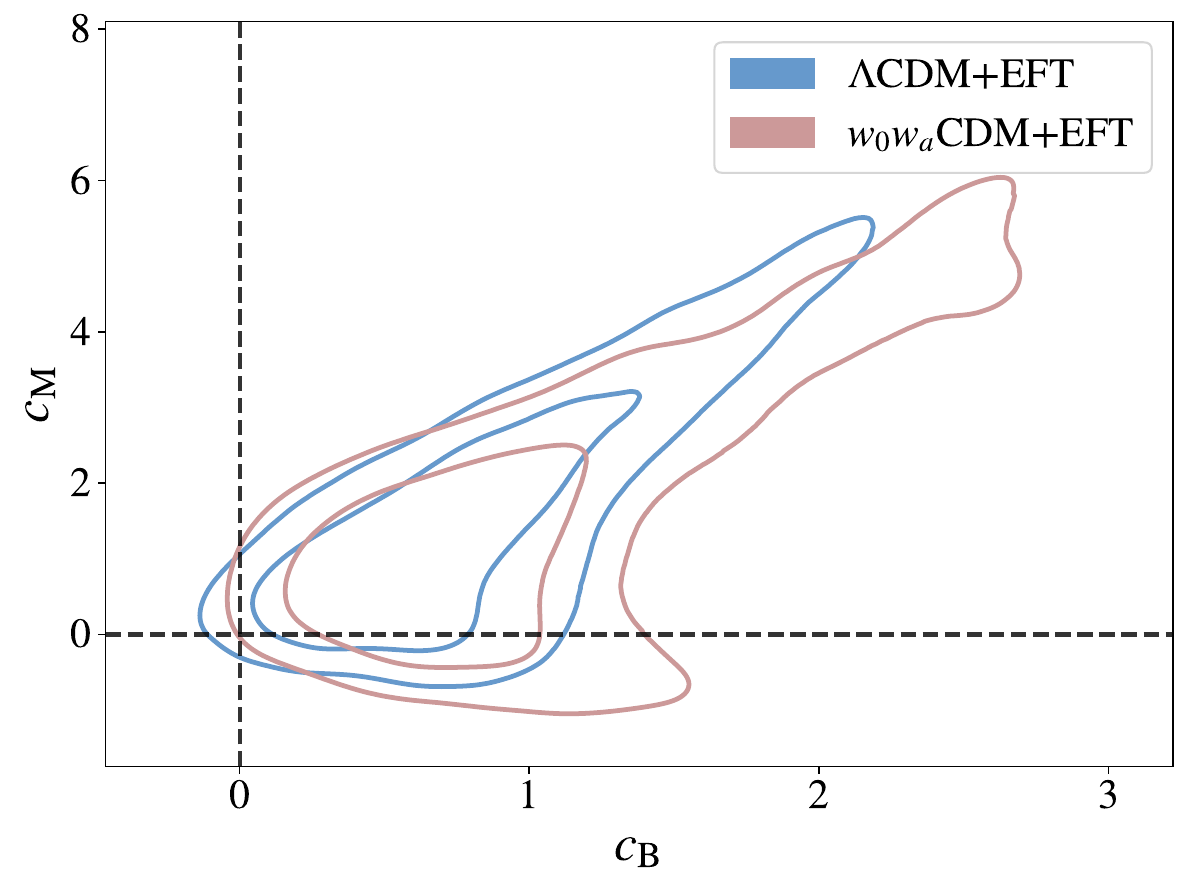}
    \end{minipage}  
    \caption{{\it{Left panel: Marginalized constraints in the $(w_0,w_a)$ plane for the
$w_0w_a$CDM model under different perturbative descriptions. The gray line
corresponds to $w_0+w_a=-1$, while the $\Lambda$CDM point
($w_0=-1$, $w_a=0$) is located at the intersection of the axes.
Right panel: Marginalized constraints in the $(c_{\rm B},c_{\rm M})$ plane for
the $\Lambda$CDM+EFT and $w_0w_a$CDM+EFT models. The contours correspond to
the 68\% and 95\% confidence regions.}}}
    \label{fig:mcmc_w0wa}
\end{figure}

The EFT analysis therefore provides an instructive comparison. Remarkably, the
neutrino-mass tension is alleviated even without introducing a dynamical
dark-energy background. This further illustrates that the inferred
neutrino-mass constraint is not a uniquely determined observable, but depends
on the physical assumptions adopted for the dark sector. We therefore present
the EFT results not as an alternative determination of the neutrino mass, but
as an explicit demonstration of the degree to which neutrino-mass constraints
remain model dependent.

In order to further understand the origin of these effects, in the left panel of
Fig.~\ref{fig:mcmc_w0wa} we present the constraints in the $(w_0,w_a)$ plane for
the $w_0w_a$CDM model under different perturbative assumptions. We observe that
allowing for dark-energy clustering modifies the preferred region of parameter
space, leading to shifts in both $w_0$ and $w_a$ compared to the smooth
dark-energy case. In particular, clustering dark energy favors parameter
combinations that effectively enhance the suppression of structure growth,
thereby reinforcing the degeneracy with neutrino mass. The displacement of the
contours away from the $\Lambda$CDM point $(w_0=-1, w_a=0)$ highlights that
the inferred neutrino-mass constraints are closely linked to the assumed
dark-energy dynamics.

Importantly, these shifts occur despite the fact that the underlying data set
remains unchanged. This demonstrates that the inferred dark-energy evolution
and the inferred neutrino mass cannot be treated as independent quantities.
Instead, both are determined simultaneously along a shared degeneracy direction
governed by the growth history of cosmic structures.

Additionally, in the right panel of Fig.~\ref{fig:mcmc_w0wa}, we show the marginalized posterior
distributions of the EFT parameters $c_{\rm M}$ and $c_{\rm B}$ for the
$\Lambda$CDM+EFT and $w_0w_a$CDM+EFT models. The region
$c_{\rm M}<0$ and $c_{\rm B}<0$ is excluded by gradient-stability
requirements. The data mildly prefer a nonzero braiding parameter
$\alpha_{\rm B}$ and a nonzero Planck-mass running parameter
$\alpha_{\rm M}$, corresponding to a departure from minimally coupled gravity.

We stress that the neutrino mass discussed above is treated here as an
effective parameter. Any indication of a negative value should not be
interpreted as evidence for negative physical neutrino masses, but rather as a
signature of underlying model degeneracies, unidentified systematics, or new
physics beyond the standard cosmological framework.

\begin{figure}[htbp]
        \centering
    \begin{minipage}{0.49\textwidth}
        \centering
        \includegraphics[width=\textwidth]{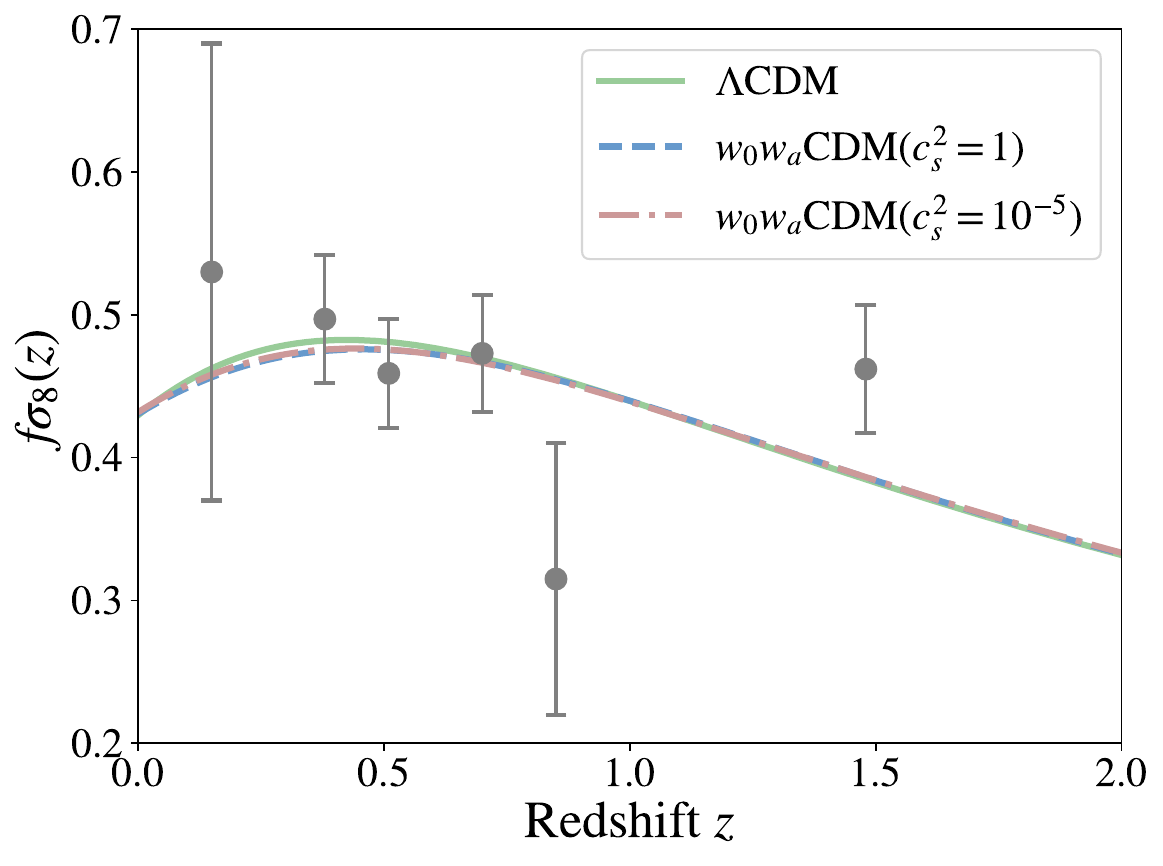}
    \end{minipage}
    \hfill
    \begin{minipage}{0.49\textwidth}
        \centering
        \includegraphics[width=\textwidth]{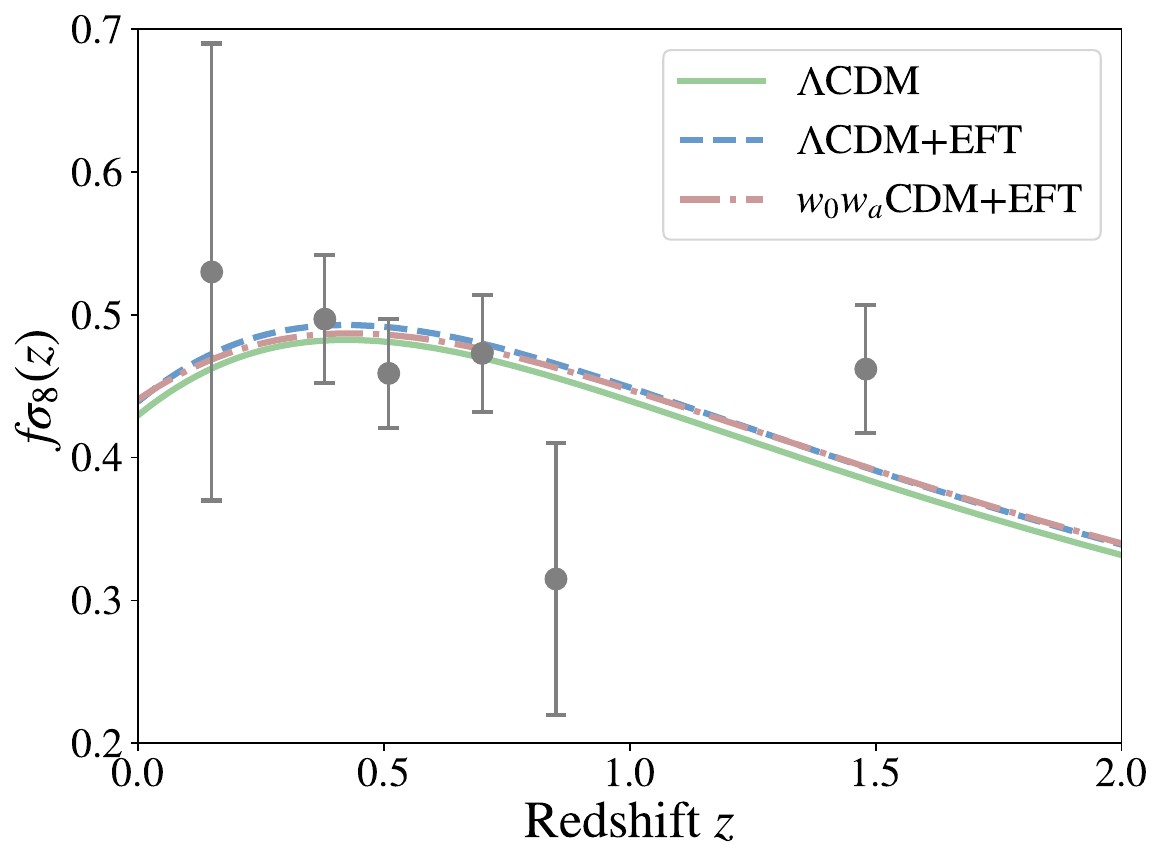}
    \end{minipage}     
    \caption{{\it{Growth-rate measurements $f\sigma_8(z)$ together with the predictions of the
best-fit cosmological models. Left panel: effective-fluid description.
Right panel: EFT description. Despite the substantial differences in the
inferred neutrino-mass constraints, the predicted growth histories remain
remarkably similar, illustrating the degeneracy between neutrino
free-streaming and dark-energy perturbations discussed in the text.}}}
    \label{fig:growth_model}
\end{figure}

\begin{figure*}[htbp]
    \centering
    \includegraphics[width=\textwidth]{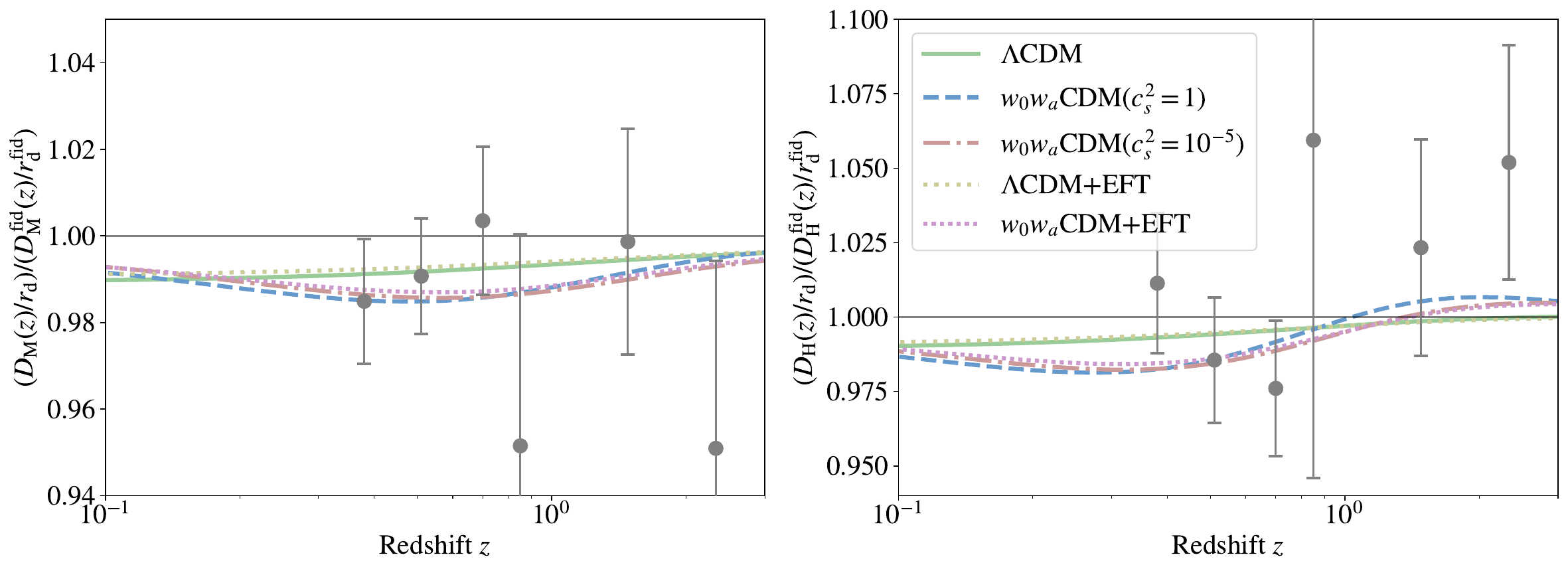}    
    \caption{{\it{Residuals of the BAO distance measurements $D_{\rm M}/r_d$ (left panel) and
$D_{\rm H}/r_d$ (right panel) relative to the \textit{Planck} 2018 best-fit
$\Lambda$CDM model. The curves show the predictions of the best-fit models
considered in this work. Despite the different inferred neutrino-mass and
dark-energy parameters, all models provide very similar BAO-distance
predictions, illustrating that current observations primarily constrain
degeneracy directions in parameter space rather than the neutrino mass
independently.}}}
    \label{fig:dm_dh_model}
\end{figure*}

In Fig.~\ref{fig:growth_model} and Fig.~\ref{fig:dm_dh_model}, we present the RSD and BAO observables, respectively, expressed as deviations from the \textit{Planck} 2018 best-fit $\Lambda$CDM model. The corresponding predictions of the cosmological scenarios considered in this work are overlaid on the observational data.
A notable feature of these figures is that, despite the substantial shifts in the inferred neutrino mass and dark-energy parameters discussed above, the predicted observables remain remarkably similar. In particular, within the effective-fluid description of the $w_0w_a$CDM model, the smooth and clustering dark-energy scenarios yield nearly indistinguishable growth-rate predictions, while exhibiting different preferred values for the underlying cosmological parameters. This behavior provides direct evidence for the degeneracy between neutrino free-streaming and dark-energy perturbations discussed in subsection \ref{subsectiondegeneracy}.

From a physical perspective, the growth suppression induced by clustering dark energy can be compensated by a corresponding shift in the effective neutrino mass, allowing different combinations of parameters to produce nearly identical predictions for structure-growth observables. At the same time, small adjustments in the background evolution help preserve agreement with the BAO measurements. Consequently, current observations constrain a combination of neutrino and dark-energy perturbation effects rather than each contribution separately.

These results therefore illustrate why neutrino-mass constraints are sensitive to assumptions regarding the perturbative properties of dark energy. The fact that models with substantially different neutrino-mass posteriors provide similarly good fits to both the RSD and BAO data demonstrates that the inferred neutrino mass is driven by an underlying degeneracy in the growth sector rather than by a direct and independent observational determination.

The corresponding quantitative constraints on the cosmological parameters are
summarized in Table~\ref{table:mcmc_result}, where we report the marginalized
posterior means and $68\%$ credible intervals for the different models
considered. While the background cosmological parameters remain remarkably
stable across the different scenarios, the inferred neutrino mass and
dark-energy parameters exhibit significant shifts. This behavior is highly
nontrivial: had the effect been driven primarily by changes in the background
expansion history, comparable shifts would also be expected in parameters such
as $\Omega_{\rm m}$ and $H_0$. Instead, the stability of the background sector
combined with the variability of the neutrino and dark-energy parameters points
directly to a perturbative origin. The dominant source of the effect is
therefore the interplay between neutrino free-streaming and dark-energy
clustering in structure-growth observables.

\begin{table*}[ht]
\centering
\caption{Summary of the cosmological parameter constraints for the models considered
in this work. Results are shown for the $\Lambda$CDM and $w_0w_a$CDM
frameworks under both the effective-fluid and effective field theory
descriptions. Quoted values correspond to the marginalized posterior means
and 68\% credible intervals.}
\begin{tabular}{lccccccc}
\hline
\multirow{2}{*}{Model} & \multirow{2}{*}{$\Omega_\mathrm{m}$} & 
\multirow{1}{*}{$H_0$} & \multirow{2}{*}{$w_0$} & \multirow{2}{*}{$w_a$} 
& \multirow{1}{*}{$\sum m_{\nu}$} &  
\multirow{2}{*}{$c_\mathrm{B}$} & \multirow{2}{*}{$c_\mathrm{M}$}\\
 &  & [$\mathrm{km}\,\mathrm{s}^{-1}\,\mathrm{Mpc}^{-1}$] &  &  & 
[$\mathrm{eV}$] &  &\\
\hline
\multirow{2}{*}{$\Lambda$CDM}
 & $0.3066$ & $67.98$ & - & - & $-0.032$  & - & - \\

 & $^{+0.0064}_{-0.0064}$ & $^{+0.49}_{-0.56}$ & - & - & $^{+0.057}_{-0.057}$  & - & - \\

\multirow{2}{*}{$w_0w_a$CDM($c_s^2=1$)}
 & $0.3124$ & $67.40$ & $-0.850$ & $-0.54$ & $-0.01$ & - & - \\

 & $^{+0.0073}_{-0.0073}$ & $^{+0.62}_{-0.62}$ & $^{+0.066}_{-0.075}$ & $^{+0.41}_{-0.32}$ & $-^{+0.11}_{-0.11}$ & - & - \\

\multirow{2}{*}{$w_0w_a$CDM($c_s^2=10^{-5}$)} 
 & $0.3110$ & $67.36$ & $-0.870$ & $-0.37$ & $-0.07$ & - & - \\
 
 & $^{+0.0075}_{-0.0075}$ & $^{+0.60}_{-0.60}$ & $^{+0.068}_{-0.068}$ & $^{+0.38}_{-0.30}$ & $^{+0.11}_{-0.11}$ & - & - \\

\multirow{2}{*}{$\Lambda$CDM+EFT}
 & $0.3081$ & $67.80$ & - & - & $0.043$ & $0.75$ & $1.44$ \\

 & $^{+0.0066}_{-0.0066}$ & $^{+0.54}_{-0.54}$ & - & - & $^{+0.072}_{-0.072}$ & $^{+0.28}_{-0.51}$ & $^{+0.64}_{-1.60}$ \\

\multirow{2}{*}{$w_0w_a$CDM+EFT}
 & $0.3106$ & $67.36$ & $-0.891$ & $-0.30$ & $-0.01$ & $0.85$ & $1.22$ \\

 & $^{+0.0072}_{-0.0072}$ & $^{+0.60}_{-0.60}$ & $^{+0.059}_{-0.070}$ & $^{+0.37}_{-0.20}$ & $^{+0.14}_{-0.14}$ & $^{+0.22}_{-0.52}$ & $^{+0.61}_{-1.44}$ \\
\hline
\end{tabular}
\label{table:mcmc_result}
\end{table*}

Finally, we complement the Bayesian analysis with a frequentist approach based
on profile likelihoods, which has recently attracted increasing attention in
cosmology
\cite{Herold:2024enb,Chebat:2025kes,Herold:2025hkb,Karwal:2024qpt}. For a
fixed value of $\sum m_\nu$, the profile likelihood is defined as
\begin{equation}
\mathcal{L}_{\rm PL}(\sum m_\nu) = \max_{\bm{\theta}, \mathcal{N}}
\mathcal{L}(\sum m_\nu, \bm{\theta}, \mathcal{N}),
\end{equation}
where $\bm{\theta}$ denotes the set of cosmological parameters and
$\mathcal{N}$ the nuisance parameters. Equivalently, we minimize
$\chi^2\equiv-2\ln\mathcal{L}$ with respect to $\bm{\theta}$ and
$\mathcal{N}$ for each fixed value of $\sum m_\nu$. Confidence intervals are
then obtained using the standard criteria $\Delta\chi^2=1$ and
$\Delta\chi^2=3.84$, corresponding respectively to the $68\%$ and $95\%$
confidence levels according to the Feldman-Cousins prescription
\cite{Feldman:1997qc}. In this frequentist analysis we focus on the effective
fluid description of dark energy, since the EFTofDE framework can also encode
modified-gravity effects and therefore involves a broader class of physical
interpretations.
 
\begin{figure}[htbp]
    \centering
    \includegraphics[width=0.6\textwidth]{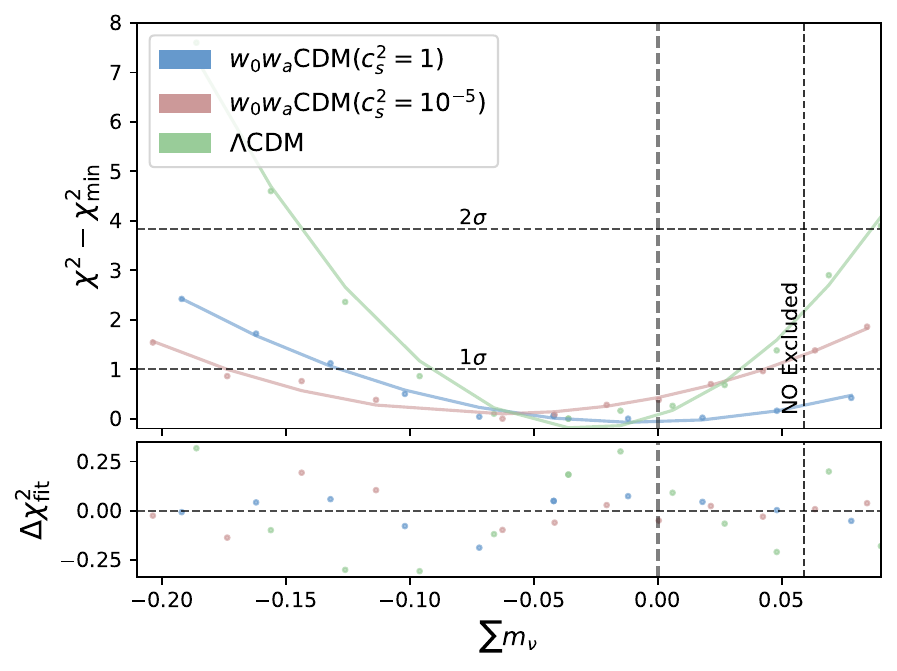}    
    \caption{{\it{Profile likelihoods for the effective neutrino mass in the $\Lambda$CDM and
$w_0w_a$CDM models, obtained from the combined BAO+CMB+SNe data set. The curves show parabolic fits to the profiled
$\Delta\chi^2$ distributions. The shaded regions indicate the $1\sigma$ and $2\sigma$
confidence intervals, while the vertical line marks the lower bound implied
by neutrino oscillation measurements for the normal ordering (NO).}}}
    \label{fig:profilelike_all}
\end{figure}

As shown in Fig.~\ref{fig:profilelike_all}, the profile-likelihood analysis
confirms the main result obtained from the Bayesian posterior. The preferred
region of $\sum m_\nu$ depends significantly on whether dark energy is treated
as smooth or clustering. In particular, allowing dark energy to cluster shifts
the profile likelihood toward smaller effective neutrino masses, in agreement
with the posterior behavior discussed above.

The agreement between the Bayesian posterior distributions and the profile
likelihoods is particularly important. Since the profile-likelihood approach is
largely insensitive to prior-volume effects, the consistency of the two methods
demonstrates that the observed shifts in the neutrino-mass constraint are not
artifacts of the Bayesian parameter volume. Instead, they correspond to genuine
changes in the likelihood structure induced by the perturbative dynamics of
dark energy.

Taken together, these results demonstrate that the commonly claimed alleviation
of the neutrino-mass tension in dynamical dark-energy models is not a robust
prediction. Rather, it depends sensitively on assumptions regarding the
perturbative behavior of dark energy. Our analysis therefore shows that
cosmological neutrino-mass constraints cannot be considered reliable unless
dark energy and neutrinos are treated consistently at the perturbation level.
Hence, the central result of this work is that neutrino-mass inference is intrinsically
model dependent in the growth sector, and that dark-energy perturbations
constitute a previously underappreciated source of uncertainty in cosmological
determinations of the neutrino mass.

\section{Conclusions}
\label{sec:concl}

The determination of the absolute neutrino-mass scale represents one of the most important goals at the interface between cosmology and particle physics. Recent cosmological analyses have produced increasingly stringent upper limits on the sum of neutrino masses, in some cases approaching or even challenging the lower bounds implied by neutrino oscillation experiments. At the same time, it has been argued that the emergence of dynamical dark-energy models may alleviate this apparent tension. In this work, we have critically reassessed the robustness of this conclusion by consistently incorporating dark-energy perturbations into the cosmological analysis.

We investigated the interplay between massive neutrinos and clustering dark energy within both an effective-fluid description and the Effective Field Theory of dark energy framework. Our analysis demonstrates that the inferred neutrino-mass constraints depend not only on the background evolution of dark energy, but also on its perturbative properties. In particular, we showed that the commonly adopted assumption of smooth dark energy can substantially affect the inferred neutrino-mass bounds.

Our central result is that the apparent alleviation of the neutrino-mass tension in dynamical dark-energy models is not a generic prediction of dynamical dark energy itself. While smooth dark energy relaxes the cosmological neutrino-mass bounds and brings them into agreement with oscillation constraints, allowing dark energy to cluster shifts the preferred neutrino mass toward smaller values and partially restores the preference already present in $\Lambda$CDM. In some cases, the preference for negative effective neutrino masses becomes even stronger. Consequently, the apparent resolution of the tension depends crucially on how dark-energy perturbations are modeled.

We traced this behavior to a physical degeneracy between neutrino free-streaming and dark-energy perturbations in structure-growth observables. Both effects modify the evolution of matter perturbations and gravitational potentials, allowing different combinations of neutrino mass and dark-energy clustering properties to generate nearly identical predictions for observables such as the matter power spectrum and the growth rate $f\sigma_8$. Importantly, we showed that significantly different neutrino-mass posteriors can provide comparably good fits to current BAO and RSD observations. This demonstrates that present cosmological data constrain a combination of neutrino and dark-energy perturbation effects rather than each component independently.

An important implication of our results is that neutrino-mass inference is intrinsically model dependent in the growth sector. Neglecting dark-energy perturbations does not merely simplify the analysis; it effectively removes part of the physically allowed parameter space and breaks a genuine cosmological degeneracy by assumption. Consequently, cosmological neutrino-mass constraints derived under the smooth-dark-energy approximation should not automatically be interpreted as robust determinations of the neutrino mass. Likewise, recent claims that dynamical dark energy resolves the neutrino-mass tension should be viewed with caution unless the perturbative properties of the dark-energy sector are treated consistently.

Furthermore, the Effective Field Theory analysis reinforces this conclusion. We found that the introduction of additional perturbative degrees of freedom can alleviate the neutrino-mass tension even in a $\Lambda$CDM background, further demonstrating that the inferred neutrino-mass bounds are sensitive to assumptions regarding the underlying dark sector. Taken together, these results indicate that the apparent preference for a particular neutrino mass is not uniquely determined by current observations, but remains closely tied to the theoretical framework adopted for dark-energy perturbations and gravity.

The novelty of this work lies in providing, for the first time, a systematic and quantitative demonstration that dark-energy perturbations constitute a significant source of uncertainty in cosmological neutrino-mass measurements. Our findings therefore identify an important and previously underappreciated limitation in current cosmological neutrino-mass inference. In this sense, the robustness of future neutrino-mass determinations cannot be established without simultaneously accounting for the perturbative dynamics of the dark-energy sector.

Finally, we mention that our analysis has been restricted to the linear regime. Since dark-energy clustering can also affect nonlinear structure formation, its impact may become even more pronounced on smaller scales. Recent numerical studies have begun exploring these effects, while upcoming surveys such as \textit{Euclid} will deliver substantially improved measurements of both the expansion and growth histories of the Universe. These observations will provide a unique opportunity to test the degeneracies identified in this work and to determine whether the current neutrino-mass tension reflects new physics, hidden systematics, or an incomplete treatment of the dark sector.

In summary, we have shown that cosmological neutrino-mass constraints are considerably less robust than often assumed. The inferred neutrino mass is not determined in isolation, but emerges from a complex interplay between neutrino free-streaming, dark-energy perturbations, and the growth of cosmic structure. A consistent treatment of all these effects is therefore essential if cosmology is to deliver reliable and model-independent measurements of the neutrino mass.

\acknowledgments

We are grateful to Qingqing Wang, Jiaxi Yu, Xin Ren and Zhiyu Lu for valuable comments. This work is supported in part by the National Key R\&D Program of China (2021YFC2203100), by the NSFC (12433002, 12261131497, 92476203), by CAS young interdisciplinary innovation team (JCTD-2022-20), by 111 Project (B23042), by CSC Innovation Talent Funds, by USTC Fellowship for International Cooperation, and by USTC Research Funds of the Double First-Class Initiative. 
ENS acknowledges the contribution of the LISA Cosmology Working Group (CosWG). They acknowledge as well support from the COST Actions CA21136 -  Addressing observational tensions in cosmology with systematics and fundamental physics (CosmoVerse) - CA23130, Bridging high and low energies in search of quantum gravity (BridgeQG) and CA21106 - COSMIC WISPers in the Dark Universe: Theory, astrophysics and  experiments (CosmicWISPers).


\appendix
\section{The detailed derivation of Eq. \eqref{eq:delta_de}}
\label{app:de_per}

We work in the conformal Newtonian gauge,
\begin{equation}
    \mathrm{d} s^2=a^2(\tau)\Big [-(1-2\Psi)\mathrm{d} \tau^2+(1+2\Phi)\delta_{ij} \mathrm{d} x^{i} \mathrm{d} x^{j}\Big ],
\end{equation}
where $\tau$ denotes the conformal time. Throughout this Appendix, primes indicate derivatives with respect to conformal time. We denote the conformal Hubble parameter by $\mathcal{H}$ and the physical Hubble parameter by $H$. For simplicity, we consider only matter and dark-energy contributions. The background Friedmann equations are then
\begin{align}
    3\mathcal{H}^2 &= 8\pi Ga^2\Big (\rho_\mathrm{m}+\rho_\mathrm{de}\Big ) ,\\
    -2\mathcal{H}'&= 8\pi G\Big( \rho_\mathrm{m}+(1+3w)\rho_\mathrm{de}\Big )=\mathcal{H}^2\Big (1+3w\Omega_\mathrm{de} \Big ),
\end{align}
where $\Omega_\mathrm{de}=\rho_{\mathrm{de}} /3\mathcal{H}^2$.

The linearized continuity and Euler equations for a fluid with equation-of-state parameter $w=p/\rho$ are
\begin{align}
    \delta' &= - \Big (1+\frac{p}{\rho} \Big ) (\theta-3\Phi') -3 \mathcal{H} \Big ( \frac{\delta p}{\delta \rho} -\frac{p}{\rho} \Big ) \delta, \\
    \theta' &= -\Big( \mathcal{H} + \frac{p'}{\rho+p} \Big ) \theta -\frac{1}{p+\rho} \Big (\nabla^2\delta p-\frac{2}{3}\nabla^4 \Pi \Big )-\nabla^2\Psi,
\end{align}
where $\delta=\delta \rho/\rho$ is the density contrast, $\theta \equiv \nabla \cdot \mathbf{v}$ is the velocity divergence, and $\Pi$ denotes the anisotropic stress. In general, the pressure perturbation is determined by the effective sound speed $c_s^2$ through
\begin{equation}
\delta p=c_s^2\,\delta\rho+3\mathcal{H}(1+w) \left(c_s^2-c_a^2\right)\rho \frac{\theta}{k^2},
\end{equation}
where $c_a^2\equiv \Dot{p}/\Dot{\rho}=w-\frac{w'}{3\mathcal{H}(1+w)}$ is the adiabatic sound speed. The perturbation equations can therefore be rewritten as
\begin{align}
    \label{eq:per_delta}
    &\delta'+3\mathcal{H}(c_s^2-w)\delta+(1+w)\Big [ 9\mathcal{H}^2(c_s^2-c_a^2)\frac{1}{k^2}+1 \Big ] \theta-3(1+w)\Phi' = 0 \\
    \label{eq:per_theta}
    &\theta'+\mathcal{H}(1-3c_s^2)\theta =k^2\Psi+\frac{2}{3(p+\rho)}k^4\Pi+\frac{k^2c_s^2}{1+w}\delta.
\end{align}
These equations are valid for arbitrary values of $w$ and $c_s^2$. In the present work we focus on fluids with vanishing anisotropic stress and therefore set $\Pi=0$. Combining the above equations yields
\begin{align}
    \label{eq:theta_expression}
    \theta &= \frac{3(1+w)\Phi'-\delta'-3\mathcal{H}(c_s^2-w)\delta}{(1+w)\Big [ 9\mathcal{H}^2(c_s^2-c_a^2)\frac{1}{k^2}+1 \Big ]}, \\
    \label{eq:theta'_expression}
    \theta' &= k^2\Psi- \mathcal{H}(1-3c_s^2)\theta+\frac{k^2c_s^2}{1+w}\delta,
\end{align}

In order to determine the evolution of $\delta$, one must additionally specify the evolution of the gravitational potentials $\Psi$ and $\Phi$. From the perturbed Einstein equations, one obtains the Poisson equation
\begin{equation}
    -k^2\Phi=4\pi Ga^2 \sum_i \rho_i \Big( \delta_i+3\mathcal{H}(1+w_i)\frac{\theta_i}{k^2} \Big ).
\end{equation}
During a purely matter-dominated epoch, $\Omega_\mathrm{m}=1$ and $\Omega_\mathrm{de}=0$. In this regime, the matter perturbation admits a growing mode, $\delta_\mathrm{m}\propto a$, and a decaying mode, $\delta_\mathrm{m}\propto a^{-3/2}$, while the scale factor evolves as $a\propto\tau^2$. The corresponding solutions are
\begin{equation}
    \delta_\mathrm{m} = \delta_0\Big( a+3\frac{H_0^2}{k^2} \Big), \quad \theta_\mathrm{m} = -\delta_0 H_0^2 a^{1/2},
\end{equation}
while the gravitational potential remains constant, namely
\begin{equation}
    \Phi = -\frac{3}{2k^2}H_0^2\delta_0.
\end{equation}

In the following we consider the clustering limit $c_s^2\simeq0$. In this case, the velocity perturbation equation for dark energy, Eq.~\eqref{eq:per_theta}, reduces to
\begin{equation}
    \theta_\mathrm{de}' +\mathcal{H}\theta_\mathrm{de}= k^2\Psi,
\end{equation}
which is identical to the corresponding matter equation. Hence,
\begin{equation}
    \theta_\mathrm{de}=\theta_\mathrm{m}=-\delta_0 H_0^2 a^{1/2}.
\end{equation}
Therefore, dark energy and matter become comoving in the limit $c_s^2\simeq0$, independently of scale (this conclusion is valid for any equation-of-state parameter $w$, provided that $c_s^2\rightarrow0$ during the matter-dominated era). Assuming in addition a constant equation of state, $w=\mathrm{const}$, Eq.~\eqref{eq:per_delta} becomes
\begin{equation}
    \begin{aligned}
&\delta_\mathrm{de}'-3\mathcal{H}w\delta_\mathrm{de}+(1+w)\Big [ -9w\frac{\mathcal{H}^2}{k^2}+1 \Big ] \Big(-\delta_0 H_0^2 a^{1/2} \Big) = 0.
\end{aligned}
\end{equation}
We consider an ansatz of the form
\begin{equation}
    \delta_\mathrm{de}=A(k)a+B(k),
\end{equation}
which, upon substitution into the above equation, yields
\begin{equation}
    \delta_\mathrm{de} = \delta_0 (1+w) \Big ( \frac{1}{1-3w}a+\frac{3H_0^2}{k^2}\Big ).
\end{equation}
Finally, in the limit $c_s^2\simeq0$ during matter domination, the dark-energy comoving density contrast is related to the matter comoving density contrast through
\begin{equation}
    \Delta_\mathrm{de}= \delta_\mathrm{de}+3\mathcal{H}(1+w)\frac{\theta_\mathrm{de}}{k^2} =\frac{1+w}{1-3w}\delta_0 a=\frac{1+w}{1-3w}\Delta_\mathrm{m},
\end{equation}
which holds on all scales.

\section{Allowing the dark-energy sound speed to vary}
\label{app:free cs2}

In the main analysis we adopted representative values of the dark-energy sound
speed in order to compare the smooth and clustering regimes. Here we relax this
assumption and treat $c_s^2$ as a free parameter. Following the
\textit{Planck} Collaboration~\cite{Planck:2015bue}, we sample
$\log_{10}c_s^2$ with a flat prior $\mathcal{U}[-10,0]$, which allows an
efficient exploration of both the smooth and strongly clustering limits.

\begin{figure*}[ht!]
    \centering
    \begin{minipage}{0.49\textwidth}
        \centering
        \includegraphics[width=\textwidth]{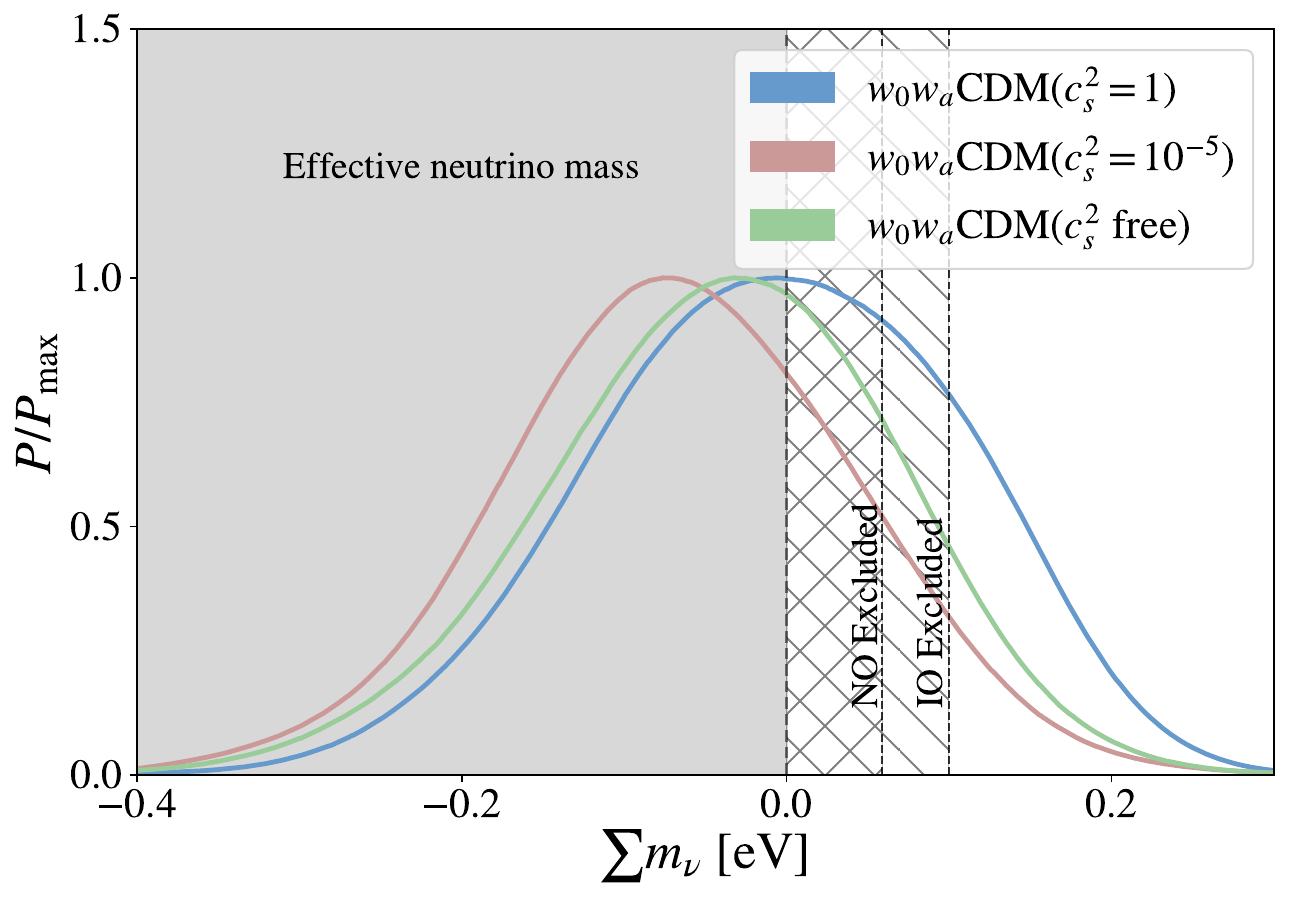}
    \end{minipage}
    \hfill
    \begin{minipage}{0.49\textwidth}
        \centering
        \includegraphics[width=\textwidth]{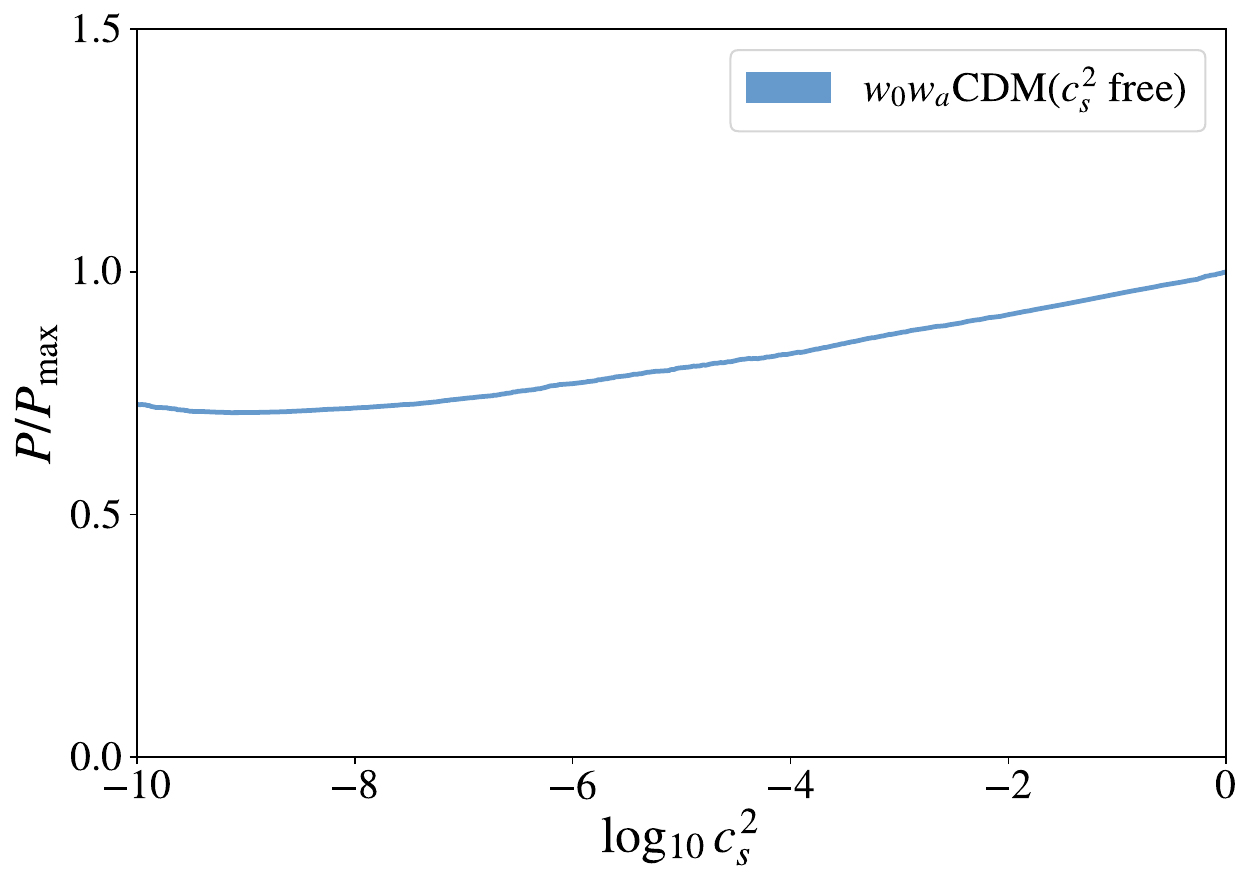}
    \end{minipage}
    \caption{\it{Left panel: Normalized one-dimensional posterior distributions of the
effective neutrino mass $\sum m_\nu$ in the $w_0w_a$CDM model when the
dark-energy sound speed is allowed to vary freely. Results are obtained from
the combined BAO+CMB+SNe data set. The lower bounds implied by neutrino
oscillation measurements for the normal ordering (NO) and inverted ordering
(IO) are also shown. Right panel: Normalized marginalized posterior
distribution of $\log_{10}c_s^2$, illustrating the weak constraints placed on
the dark-energy sound speed by current observations.}}
    \label{fig:app_free_cs2}
\end{figure*}

The resulting constraints are shown in
Fig.~\ref{fig:app_free_cs2}. As expected, current cosmological observations
provide only weak constraints on the dark-energy sound speed, with the
posterior remaining broad over most of the prior range.

To quantify the parameter degeneracies, we compute the correlation matrix from
the covariance matrix $C_{\alpha\beta}$
\begin{equation}
R_{\alpha\beta}\equiv
\frac{C_{\alpha\beta}}
{\sqrt{C_{\alpha\alpha}C_{\beta\beta}}},
\end{equation}
for the parameter set
$\{w_0,w_a,\log_{10}c_s^2,\Omega_{\mathrm m},\sum m_{\nu} \}$.
The resulting correlation matrix is
\begin{equation}
R\Big \{w_0, w_a, \log_{10}c_s^2, \Omega_{\mathrm{m}}, \sum m_\nu \Big \} = 
\begin{pmatrix}
1.000 & -0.895 &  0.139 &  0.471 &  0.449 \\
-0.895 & 1.000 & -0.180 & -0.492 & -0.725 \\
 0.139 & -0.180 & 1.000 &  0.099 &  0.179 \\
 0.471 & -0.492 &  0.099 & 1.000 &  0.603 \\
 0.449 & -0.725 &  0.179 &  0.603 & 1.000
\end{pmatrix}.
\end{equation}

The strongest correlations involving the neutrino mass are those with $w_a$
($R=-0.725$), $\Omega_{\mathrm m}$ ($R=0.603$), and $w_0$ ($R=0.449$),
confirming that neutrino-mass constraints remain closely connected to the
dark-energy sector. In contrast, the correlation between $\sum m_\nu$ and
$\log_{10}c_s^2$ is weak ($R=0.179$). This indicates that the sound speed
itself is not the dominant source of uncertainty in the neutrino-mass
determination.

In summary, allowing $c_s^2$ to vary freely does not alter the main conclusions
of this work. The dominant degeneracies involve the dark-energy equation of
state and its impact on structure growth, while the sound speed remains only
weakly constrained by current observations.

\section{Separate effects of effective neutrino mass and dark-energy perturbations}
\label{app:separate_effects}

In this Appendix we illustrate separately the effects of the effective neutrino mass and dark-energy perturbations on the cosmological observables relevant for our analysis. The purpose is to provide a direct visualization of the degeneracy discussed in Sec.~\ref{subsectiondegeneracy}.

\begin{figure*}[ht!]
\centering
\includegraphics[width=\textwidth]{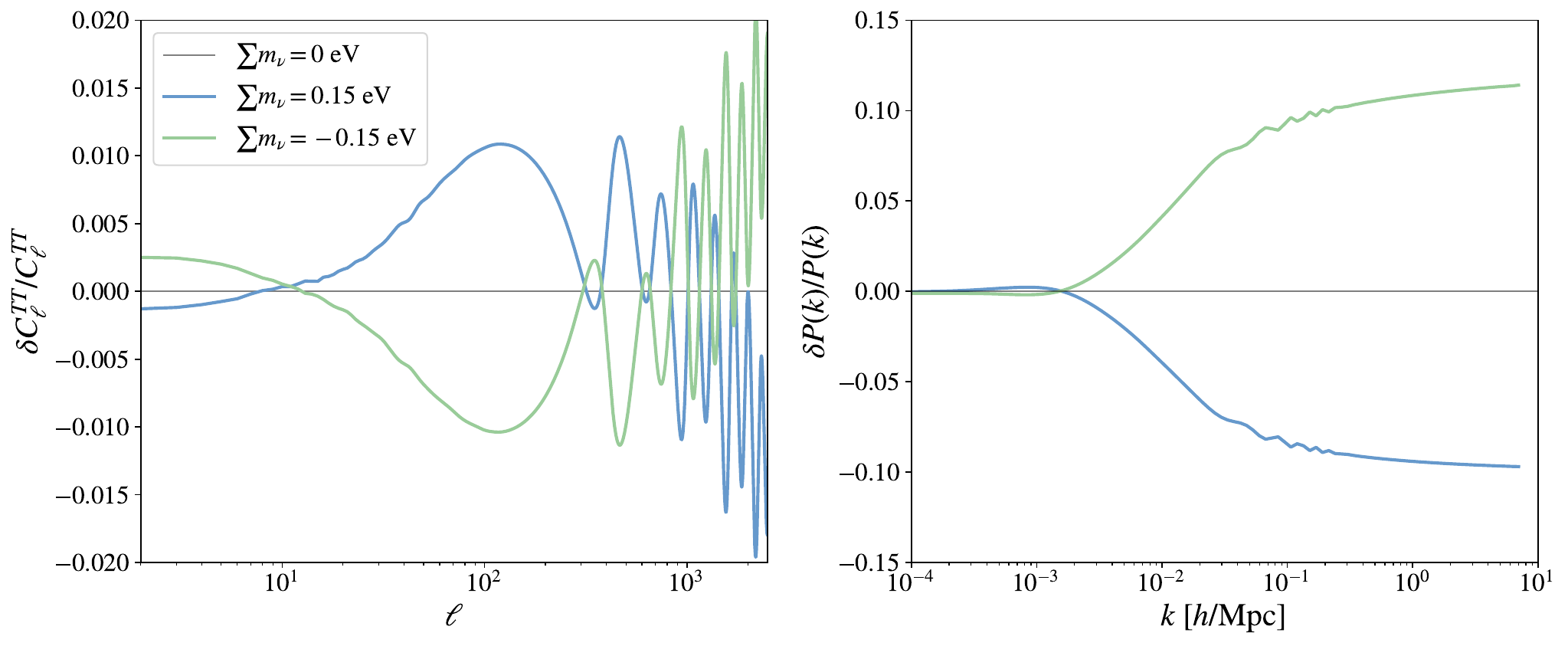}
\caption{\it{Residuals of the CMB temperature power spectrum (left panel) and the matter
power spectrum at $z=0$ (right panel) in the $\Lambda$CDM model for different
values of the effective neutrino mass. The black solid line corresponds to the
massless-neutrino case, while the blue and green curves correspond to
$\sum m_\nu=0.15\,\mathrm{eV}$ and
$\sum m_\nu=-0.15\,\mathrm{eV}$, respectively.}}
\label{fig:cmb_mv_eff}
\end{figure*}

Figure~\ref{fig:cmb_mv_eff} shows the residuals of the CMB temperature power spectrum and the matter power spectrum at $z=0$ for representative positive and negative values of the effective neutrino mass within the $\Lambda$CDM framework. As expected, positive effective neutrino masses suppress the matter power spectrum on scales below the neutrino free-streaming scale, while negative effective masses enhance structure growth. The impact on the CMB temperature spectrum is comparatively smaller.

\begin{figure*}[htbp]
\centering
\includegraphics[width=\textwidth]{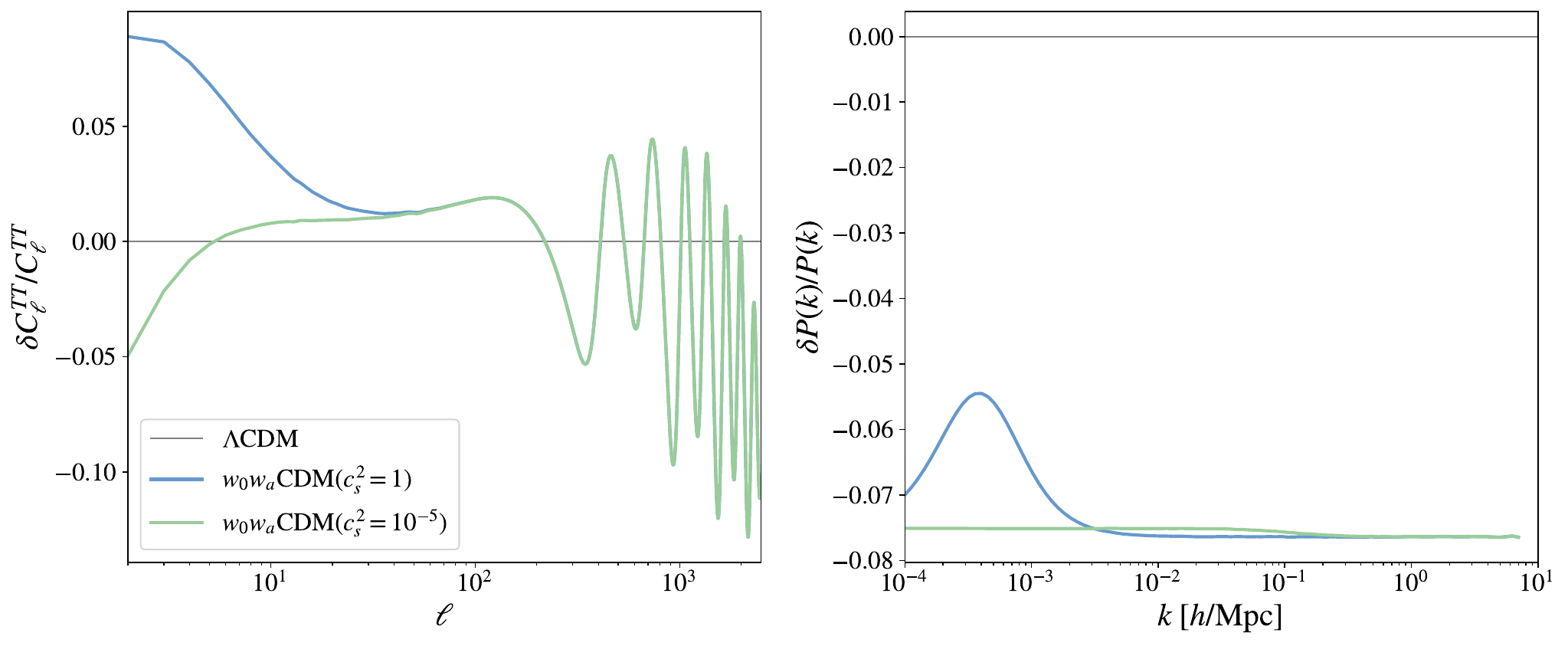}
\caption{\it{Residuals of the CMB temperature power spectrum (left panel) and the matter
power spectrum at $z=0$ (right panel) illustrating the impact of dark-energy
perturbations. The black solid line corresponds to the $\Lambda$CDM model,
while the blue and green curves correspond to the $w_0w_a$CDM model with
smooth ($c_s^2=1$) and clustering ($c_s^2=10^{-5}$) dark energy,
respectively. For illustration, we adopt $w_0=-0.7$ and $w_a=-0.5$.}}
\label{fig:cmb_de_eff}
\end{figure*}

Figure~\ref{fig:cmb_de_eff} shows the corresponding residuals for smooth dark energy ($c_s^2=1$) and clustering dark energy ($c_s^2=10^{-5}$) in the $w_0w_a$CDM framework. While the CMB temperature spectrum is only mildly affected, dark-energy clustering produces a noticeable modification of the matter power spectrum through its impact on the evolution of gravitational potentials and matter perturbations.

Comparing the two figures reveals the origin of the degeneracy. Although neutrino free-streaming and dark-energy clustering arise from different physical mechanisms, both affect the observables used to constrain structure growth. Consequently, changes in the neutrino sector can be partially compensated by changes in the perturbative behavior of dark energy, leading to similar predictions for the matter power spectrum and growth observables. This degeneracy is responsible for the shifts in the inferred neutrino-mass constraints discussed in Sec.~\ref{dataresults}.

 \section{Results for $w$CDM model}
\label{app:wcdm}

\begin{figure*}[ht!]
    \centering
    \begin{minipage}{0.49\textwidth}
        \centering
        \includegraphics[width=\textwidth]{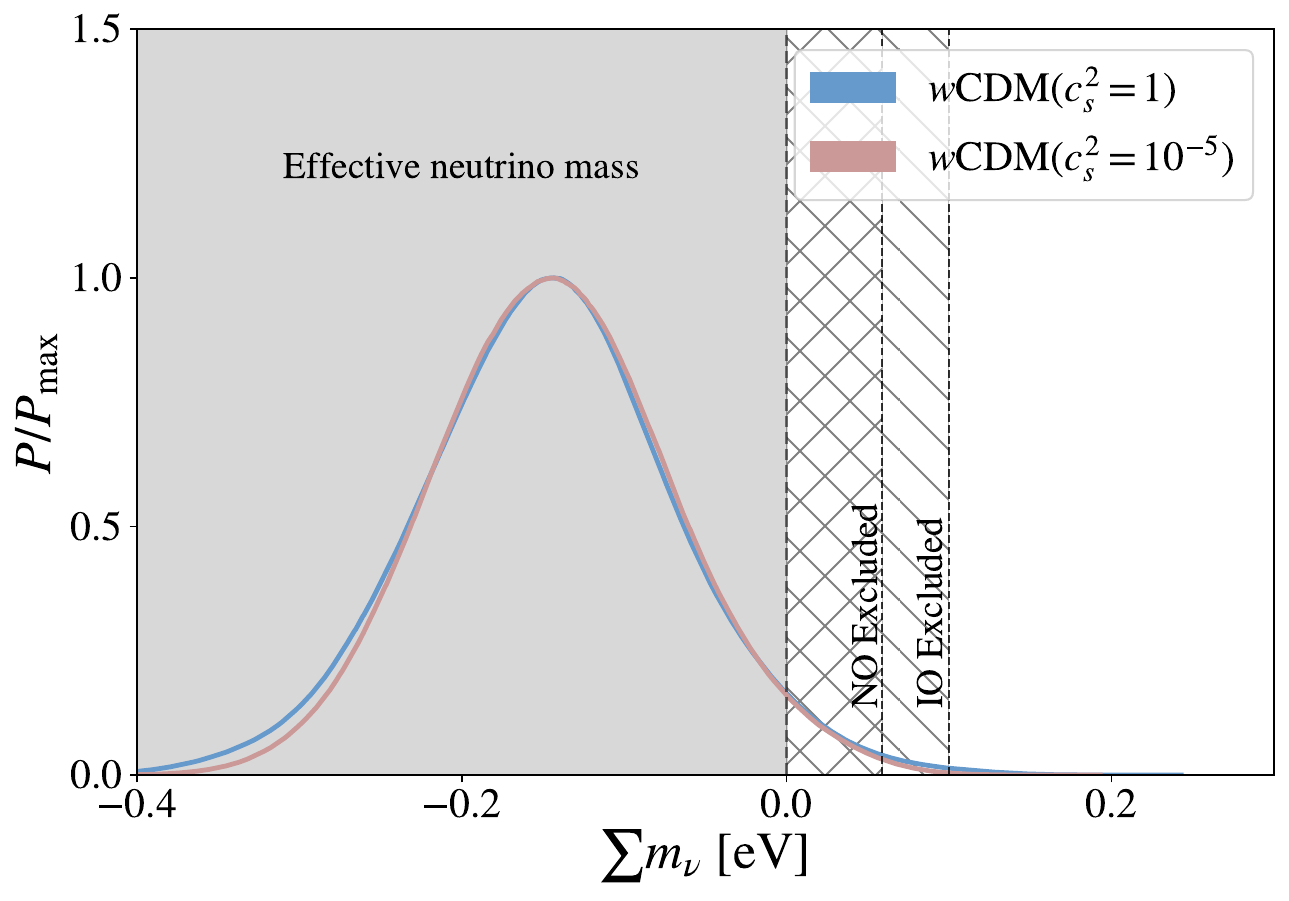}
    \end{minipage}
    \hfill
    \begin{minipage}{0.49\textwidth}
        \centering
        \includegraphics[width=\textwidth]{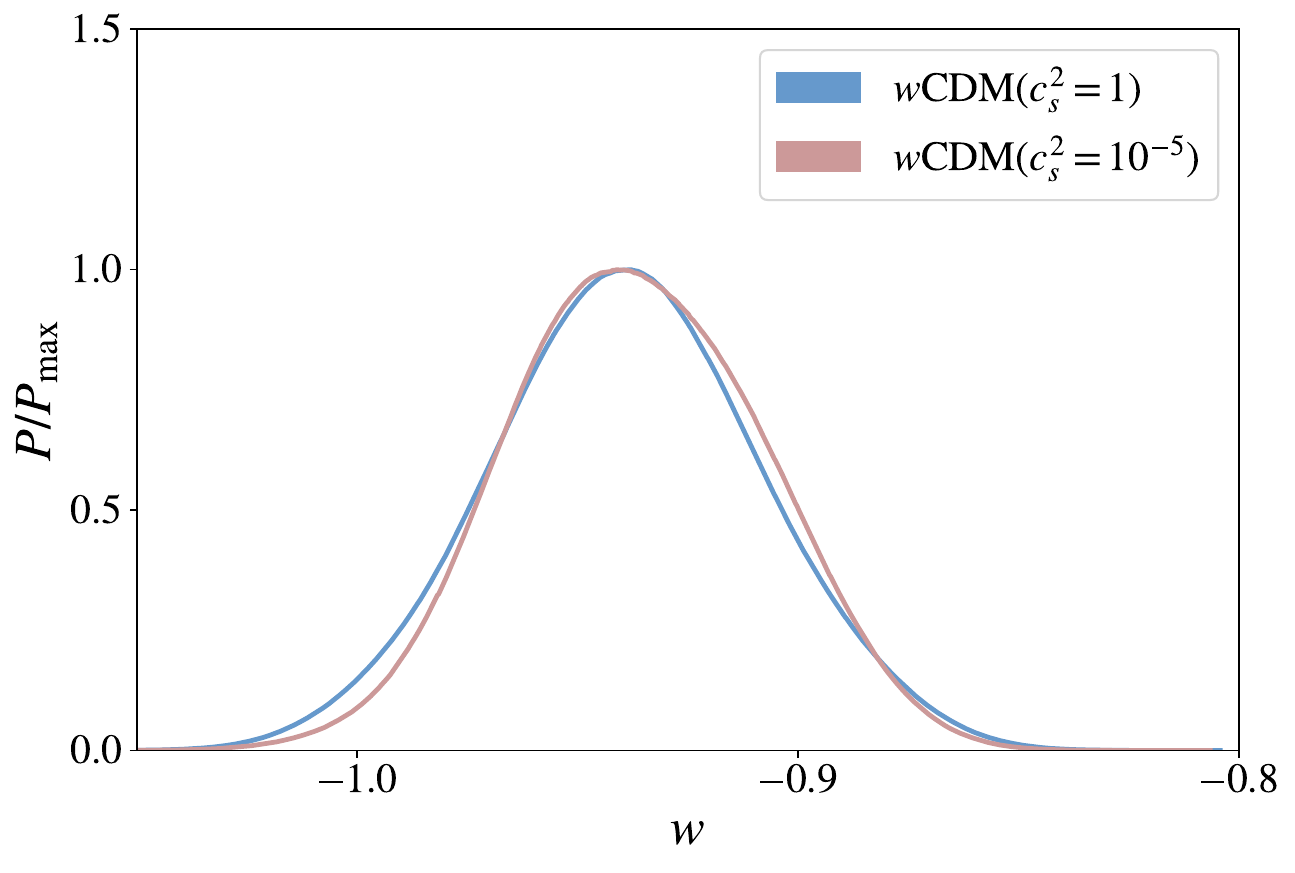}
    \end{minipage}
    \caption{\it{Left panel: Normalized one-dimensional posterior distributions of the
effective neutrino mass $\sum m_\nu$ in the $w$CDM model, obtained from the
combined BAO+CMB+SNe data set. The lower bounds implied by neutrino
oscillation measurements for the normal ordering (NO) and inverted ordering
(IO) are also shown. Right panel: Normalized marginalized posterior
distribution of the equation-of-state parameter $w$. The preference for
values close to $w=-1$ explains the limited impact of dark-energy clustering
on the inferred neutrino-mass constraint in the $w$CDM framework.}}
    \label{fig:app_wcdm}
\end{figure*}

For completeness, Fig.~\ref{fig:app_wcdm} presents the constraints obtained for
the $w$CDM model. In contrast to the $w_0w_a$CDM case discussed in the main
text, allowing dark energy to cluster produces only a minor impact on the
inferred neutrino-mass constraint. The reason is that the data constrain the
equation-of-state parameter to remain close to the cosmological-constant value,
$w\simeq -1$. As follows from Eq.~(\ref{eq:delta_de}), the
amplitude of dark-energy perturbations is proportional to $(1+w)$ and therefore
becomes strongly suppressed as $w\rightarrow -1$, independently of the precise
value of the sound speed.

Consequently, even in the clustering regime ($c_s^2\ll 1$), dark-energy
perturbations remain too small to generate a significant degeneracy with the
effective neutrino mass. The resulting neutrino-mass constraints therefore
remain close to those obtained in the smooth-dark-energy limit. This explains
why the $w$CDM model does not exhibit the substantial shifts observed in the
$w_0w_a$CDM framework and is not included among the main results of the paper.


\bibliographystyle{JHEP}
\bibliography{ref.bib}

\end{document}